\documentclass[useAMS,usenatbib]{mn2e}

\usepackage{revsymb}
\usepackage{amsmath}
\usepackage{amsfonts}
\usepackage{amssymb}
\usepackage{graphicx}

\title[CMB reconstruction ]
  {Can one reconstruct masked CMB sky?}
\author[R. Aurich and S.~Lustig]
  {R.~Aurich and S.~Lustig \\
  Institut f\"ur Theoretische Physik, Universit\"at Ulm,\\
  Albert-Einstein-Allee 11,\\ D-89069 Ulm, Germany
}

\date{}

\pagerange{\pageref{firstpage}--\pageref{lastpage}} \pubyear{2010}

\def\LaTeX{L\kern-.36em\raise.3ex\hbox{a}\kern-.15em
    T\kern-.1667em\lower.7ex\hbox{E}\kern-.125emX}

\begin{document}

\def\bfis{\hbox{\scriptsize\rm i}}
\def\bfi{\hbox{\rm i}}
\def\bfj{\hbox{\rm j}}

\newcommand{\apj}{{Astrophys.\ J. }}
\newcommand{\apjs}{{Astrophys.\ J.\ Supp. }}
\newcommand{\apjl}{{Astrophys.\ J.\ Lett. }}
\newcommand{\aj}{{Astron.\ J. }}
\newcommand{\prl}{{Phys.\ Rev.\ Lett. }}
\newcommand{\prd}{{Phys.\ Rev.\ D }}
\newcommand{\mnras}{{Mon.\ Not.\ R.\ Astron.\ Soc. }}
\newcommand{\araa}{{ARA\&A }}
\newcommand{\aap}{{Astron.\ \& Astrophy. }}
\newcommand{\nat}{{Nature }}
\newcommand{\cqg}{{Class.\ Quantum Grav.\ }}

\setlength{\topmargin}{-1cm}

\label{firstpage}

\maketitle

\begin{abstract}
The CMB maps obtained by observations always possess domains
which have to be masked due to severe uncertainties with respect to
the genuine CMB signal.
Cosmological analyses ideally use full CMB maps in order to get
e.\,g.\ the angular power spectrum.
There are attempts to reconstruct the masked regions at least
at low resolutions, i.\,e.\ at large angular scales,
before a further analysis follows.
In this paper, the quality of the reconstruction is investigated
for the ILC (7yr) map as well as for 1000 CMB simulations of the
$\Lambda$CDM concordance model.
The latter allows an error estimation for the reconstruction algorithm
which reveals some drawbacks.
The analysis points to errors of the order of a significant fraction
of the mean temperature fluctuation of the CMB.
The temperature 2-point correlation function $C(\vartheta)$ is
evaluated for different reconstructed sky maps
which leads to the conclusion that it is safest to compute it on the cut-sky.
\end{abstract}



\section{Introduction}

The cosmic microwave background (CMB) provides one of the
cornerstones of the cosmological concordance model.
The statistical properties of  our cosmological models have to
match those of the CMB in order to give an admissible model.
Thus, it is of utmost importance to reliably extract the
statistical properties of the CMB.
A main obstacle is the foreground emission of our galaxy
and of other sources
which restrict the area of the sky available for a sufficiently
clean CMB signal,
i.\,e.\ the full sky CMB signal has to be masked.
One path of statistical analysis leads to the Fourier space
in which the CMB is decomposed with respect to spherical harmonics
$Y_{lm}(\hat n)$ where the masked sky leads to a coupling
between the Fourier modes since no full sky CMB is available.
In this paper, we do not delve into these difficulties,
where upon an extensive literature exists,
but instead follow the alternative path
which allows an analysis directly in the pixel space,
which in turn is more adapted to a masked sky.
This analysis is based on the temperature two-point correlation function
$C(\vartheta)$, which is defined as
\begin{equation}
\label{Eq:C_theta}
C(\vartheta) \; := \; \left< \delta T(\hat n) \delta T(\hat n')\right>
\hspace{10pt} \hbox{with} \hspace{10pt}
\hat n \cdot \hat n' = \cos\vartheta
\hspace{10pt} ,
\end{equation}
where $\delta T(\hat n)$ is the temperature fluctuation in
the direction of the unit vector $\hat n$.
The most direct way to deal with a mask is just to use only
those pixels which are outside the mask.
In this way it was discovered by the COBE team \citep{Hinshaw_et_al_1996}
that the correlation function $C(\vartheta)$ possesses surprisingly
low power at large angles $\vartheta \gtrsim 60^\circ$.

A surprising observation is made by using the ILC map,
which represents a full sky CMB map obtained by the WMAP team
\citep{Gold_et_al_2010}.
Computing the correlation function $C(\vartheta)$
using a mask leads to a correlation function having very low power
at large scales,
whereas using the complete ILC map leads to a correlation function
which possesses higher large scale power being compatible with
the concordance model
\citep{Spergel_et_al_2003}.
One has to decide which correlation function $C(\vartheta)$
corresponds to the true CMB sky:
the one which is based on the safe pixels,
i.\,e.\ those outside the mask,
or the other one,
which would imply that most of the large scale power is generated
by those areas hidden by the galaxy,
i.\,e.\ by those pixels which have experienced much larger corrections.
This question has recently stimulated much discussions, e.\,g.\
\citep{Copi_Huterer_Schwarz_Starkman_2006,Copi_Huterer_Schwarz_Starkman_2008,%
Copi_Huterer_Schwarz_Starkman_2010,Hajian_2007,%
Aurich_Janzer_Lustig_Steiner_2007,Aurich_Lustig_Steiner_2009,%
Sarkar_Huterer_Copi_Starkman_Schwarz_2010,%
Bennett_et_al_2010,%
Efstathiou_Ma_Hanson_2009,Pontzen_Peiris_2010}.

\cite{Efstathiou_Ma_Hanson_2009} emphasise
that one has to start with the cut sky,
but before the correlation function $C(\vartheta)$ is computed
the low-order multipoles have to be reconstructed.
In this way, a stable result is obtained as long as the mask
is not too large.
The obtained correlation function is then the one with large
power at large scales.
For the details of the method see
\cite{deOliveira-Costa_Tegmark_2006,Bielewicz_Gorski_Banday_2004}.
Here we summarise only the most important ingredients.
The data vector $\vec x$ containing only the pixel values outside the mask
is related to the spherical harmonic coefficients $a_{lm}$
represented as $\vec a$ by
\begin{equation}
\label{Eq:x_as_a}
\vec x \; = \; Y \, \vec a \, + \, \vec n
\hspace{10pt} ,
\end{equation}
where $Y_{ij}$ denotes the corresponding values of $Y_{l_jm_j}(\hat n_i)$
and $\vec n$ the noise.
In the low-order multipole reconstruction, only the multipoles
with $l \leq l_{\hbox{\scriptsize max}} = 10\dots20$ are taken into account.
The methods of reconstruction differ by the choice
of a square matrix $A$
which determines the reconstructed $\vec a\,^r$ by
\begin{equation}
\label{Eq:ar_by_A}
\vec a\,^r \; = \; (Y^T A Y)^{-1} \, Y^T A \vec x
\hspace{10pt} .
\end{equation}
Setting the matrix $A$ equal to the unit matrix leads to the method
of ``direct inversion''.
To take the correlations between the pixels into account,
one can choose the covariance matrix $A^{-1}= <\vec x \cdot \vec x\,^T>$
for a reconstruction up to $l_{\hbox{\scriptsize max}}$,
which leads to the method used by \cite{deOliveira-Costa_Tegmark_2006} with
\begin{equation}
\label{Eq:covariance_matrix}
A_{ij}^{-1} \; = \;
\sum_{l=l_{\hbox{\scriptsize max}}+1}^{l_{\hbox{\scriptsize cut}}}
\frac{2l+1}{4\pi} \, P_l(\hat n_i \cdot \hat n_j) \, C_l
\end{equation}
ignoring the noise contribution.
The sum in (\ref{Eq:covariance_matrix}) runs only over those
multipole moments $C_l$, $l>l_{\hbox{\scriptsize max}}$,
that are not to be reconstructed.

The reconstructed sky map up to $l_{\hbox{\scriptsize max}}$
is then obtained from the coefficients $\vec a\,^r$.
\cite{Efstathiou_Ma_Hanson_2009} compute
the correlation function $C(\vartheta)$
from such reconstructed maps for $l_{\hbox{\scriptsize max}}=5$
to $l_{\hbox{\scriptsize max}}=20$ by using the KQ85 or the
KQ75 mask provided by the WMAP team, and it is found
that these $C(\vartheta)$ only show a negligible variation.
This is interpreted as a sign for a stable method to compute the
temperature correlation in the presence of masks.
It should be emphasised that the WMAP team bases its
7 year investigation with respect to CMB anomalies \citep{Bennett_et_al_2010}
on the method of \citep{Efstathiou_Ma_Hanson_2009}.
This leads to the conclusion
that the correlation function agrees well with the
$\Lambda$CDM concordance model and displays thus no anomalous behaviour.

The system of equations (\ref{Eq:ar_by_A}) is over-determined,
since there are much more pixel values than multipole moments with
$l \leq l_{\hbox{\scriptsize max}}$ as long as the mask and
$l_{\hbox{\scriptsize max}}$ are not too large.
Thus, there is the hope that the pixels outside the mask already
determine the low-order multipoles.
The caveat in the demonstration of \citep{Efstathiou_Ma_Hanson_2009} is,
however,
that a large Gaussian smoothing of $10^\circ$ is applied to the maps 
{\it before} the cut is applied.
Due to this smoothing there is a transfer of information from pixel values
within the mask to those outside the mask.
Since the system of equations (\ref{Eq:ar_by_A}) is already over-determined,
this additional information is readily extracted and
reveals the temperature structure within the mask
after the ``reconstruction''.
It should be noted that even a downgrade in the HEALPix resolution
can lead to an information transfer by carrying out the downgrade
before the mask is applied.

\begin{figure}
\begin{center}
{
\begin{minipage}{11cm}
\hspace*{-20pt}\includegraphics[width=9.0cm]{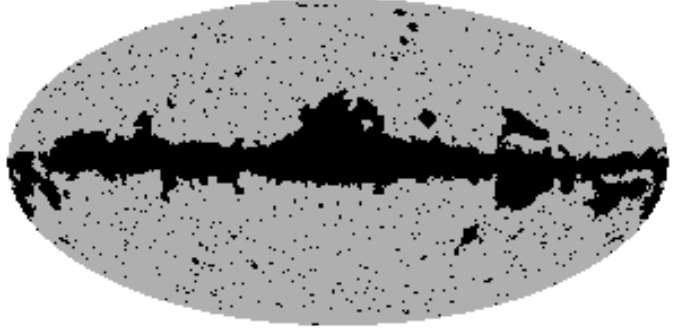}
\end{minipage}
\put(-315,67){(a) KQ85 (7yr) mask, $N_{\hbox{\scriptsize side}} = 512$}
}
\vspace*{5pt}
{
\begin{minipage}{11cm}
\hspace*{-20pt}\includegraphics[width=9.0cm]{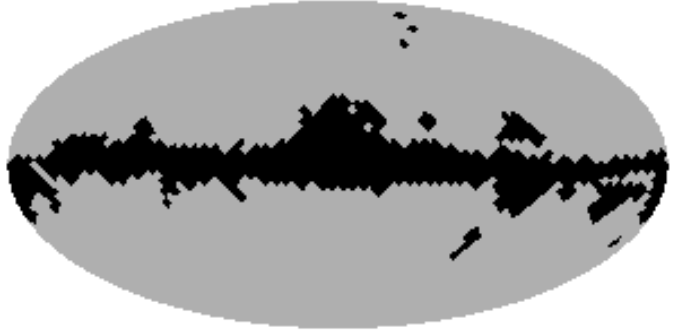}
\end{minipage}
\put(-315,67){(b) KQ85 (7yr) mask, $N_{\hbox{\scriptsize side}} = 16$, $x_{\hbox{\scriptsize th}}=0.5$ }
}
\end{center}
\caption{\label{Fig:KQ85_masks_nside_512_and_nside_16}
The KQ85 mask of the WMAP 7 year data with a pixel resolution of
$N_{\hbox{\scriptsize side}} = 512$ is displayed in panel 
\ref{Fig:KQ85_masks_nside_512_and_nside_16}a. 
The masked region is pictured in black. 
Figure \ref{Fig:KQ85_masks_nside_512_and_nside_16}b shows
the KQ85 mask downgraded to a pixel resolution of
$N_{\hbox{\scriptsize side}} =16$.
All pixels with $x(i)\le 0.5$, equation (\ref{Eq:not_masked_pixel_downgr}),
are considered as masked.
}
\end{figure}

The two masks used in this paper are the KQ85 (7yr),
shown in figure \ref{Fig:KQ85_masks_nside_512_and_nside_16}a,
and the KQ75 (7yr) mask
which are available at the LAMBDA website.
These are stored in the HEALPix
\citep{Gorski_Hivon_Banday_Wandelt_Hansen_Reinecke_Bartelmann_2005}
format with a HEALPix resolution of
$N_{\hbox{\scriptsize side}} = 512$.
The reconstruction algorithm requires masks in lower resolutions
of $N_{\hbox{\scriptsize side}} = 16$ or 32,
and a downgrade has to be carried out.
A single downgraded pixel $i$ contains $N_{\hbox{\scriptsize total}}$
pixels of the higher resolution map and from these only
$N_{\hbox{\scriptsize nm}}$
pixel values are used in the averaging process,
i.\,e.\ those that are {\it not} masked in the higher resolution map,
in order to compute the value of the downgraded pixel.
If the ratio
\begin{equation}
\label{Eq:not_masked_pixel_downgr}
x(i) \; = \;
\frac{N_{\hbox{\scriptsize nm}}}
{N_{\hbox{\scriptsize total}}}
\end{equation}
is larger than a given mask threshold
$x_{\hbox{\scriptsize th}}\in(0.0, 1.0)$,
i.\,e.\ $x(i)>x_{\hbox{\scriptsize th}}$, 
the resulting pixel is {\it not} masked, and otherwise it is masked.
The result of the KQ85 (7yr) mask for $N_{\hbox{\scriptsize side}} =16$ 
and $x_{\hbox{\scriptsize th}}=0.5$ is shown in
figure \ref{Fig:KQ85_masks_nside_512_and_nside_16}b. 
It is obvious that the size of the masked domains in masks with 
$N_{\hbox{\scriptsize side}} <512$ depends on this mask threshold
$x_{\hbox{\scriptsize th}}$.

\begin{figure}
\begin{center}
{
\begin{minipage}{11cm}
\hspace*{-20pt}\includegraphics[width=9.0cm]{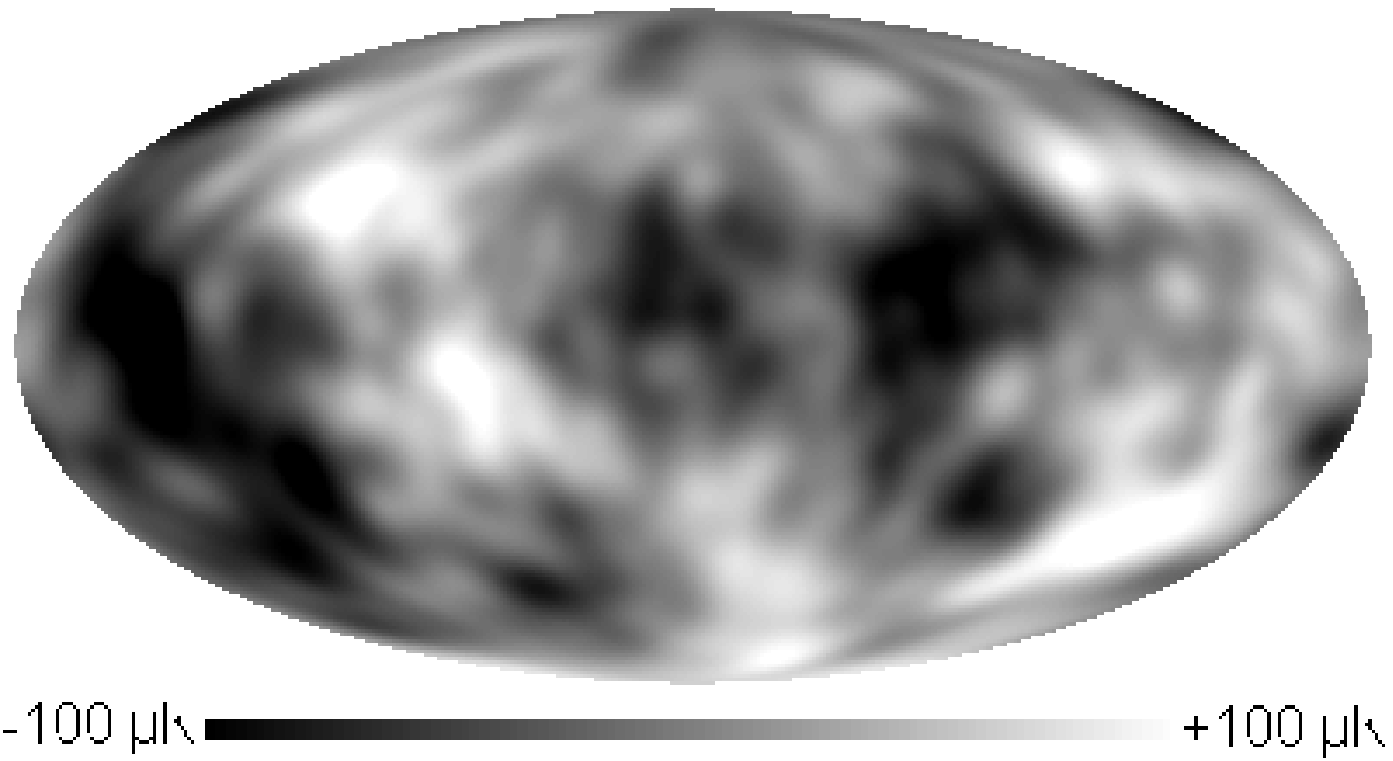}
\end{minipage}
\put(-325,75){(a) $\Lambda$CDM simulation A}
}
\vspace*{10pt}
{
\begin{minipage}{11cm}
\hspace*{-20pt}\includegraphics[width=9.0cm]{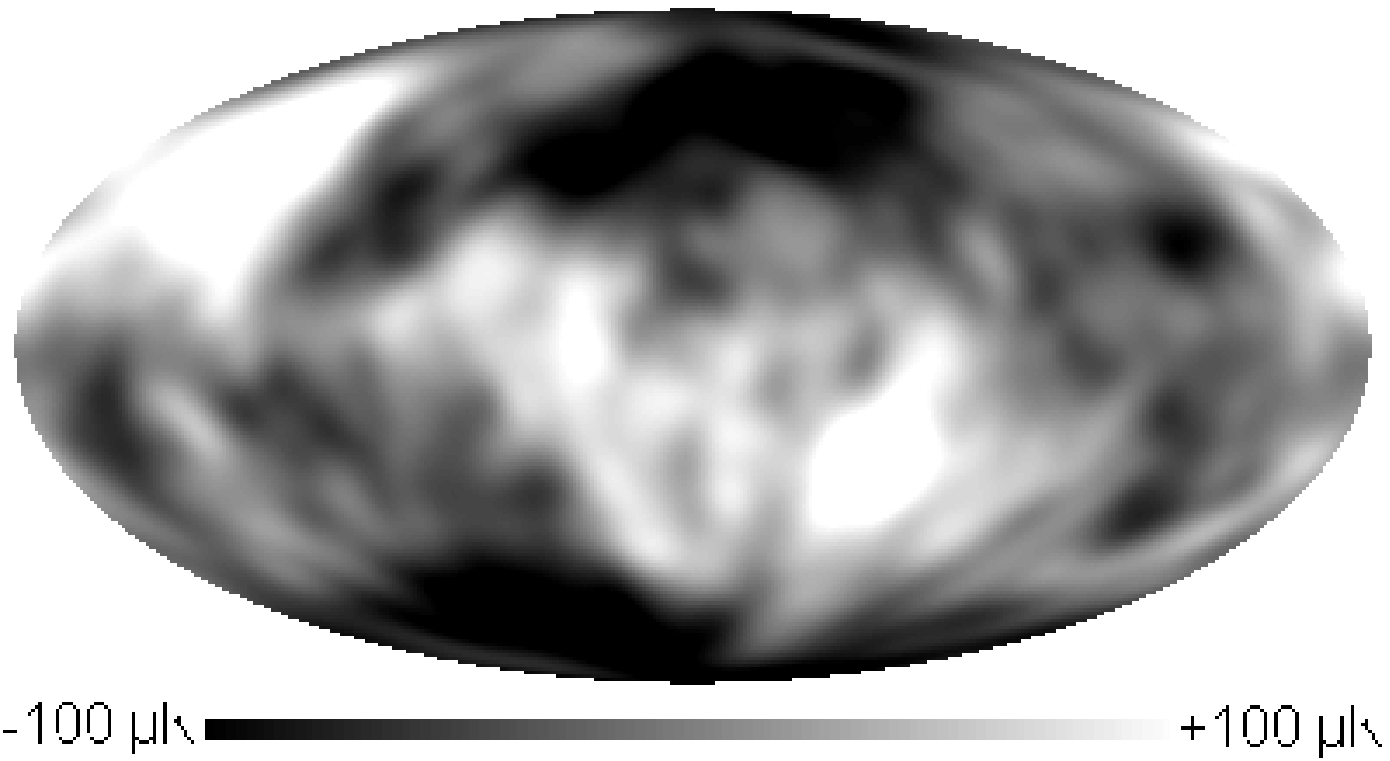}
\end{minipage}
\put(-325,75){(b) $\Lambda$CDM  simulation B}
}
\vspace*{10pt}
{
\begin{minipage}{11cm}
\hspace*{-20pt}\includegraphics[width=9.0cm]{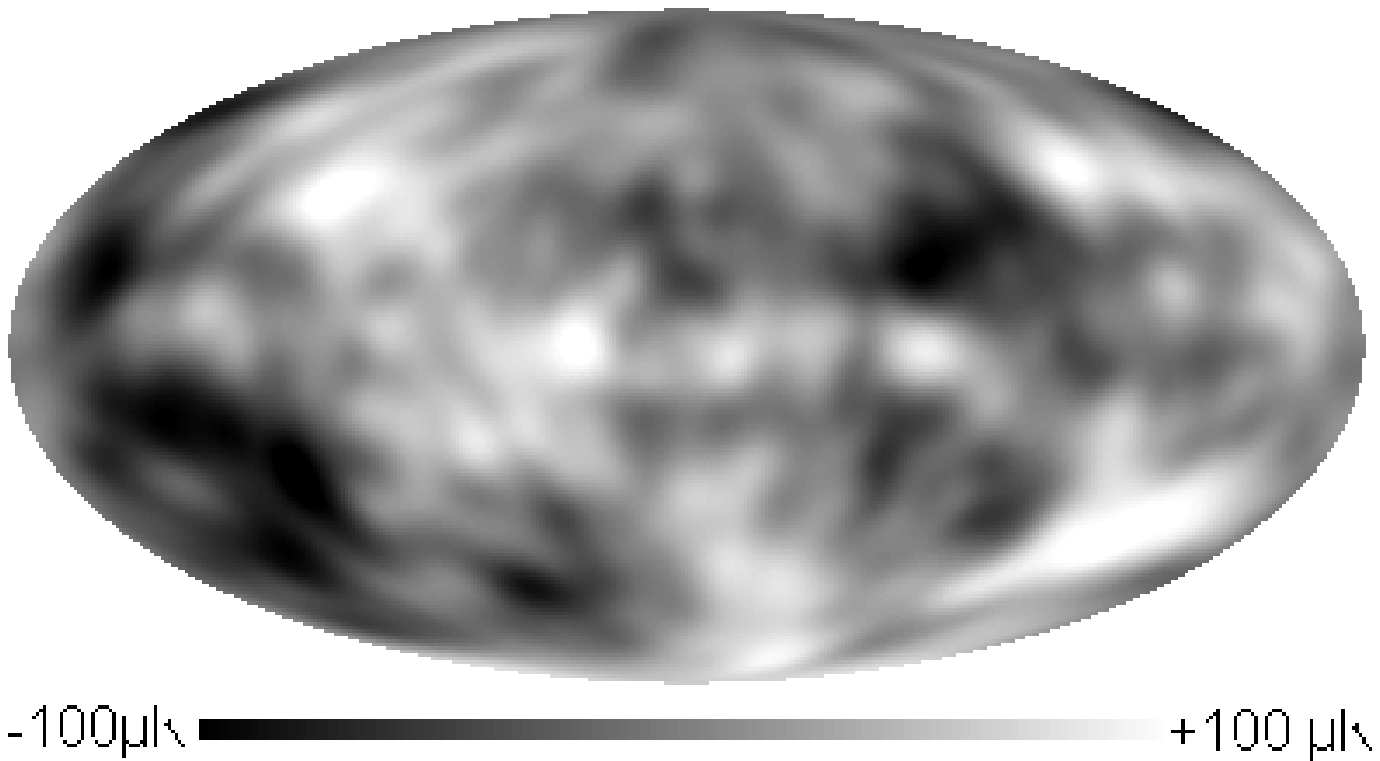}
\end{minipage}
\put(-325,75){(c) $\Lambda$CDM simulation A outside and B inside KQ85 mask}
}
\end{center}
\caption{\label{Fig:CutSky_Info_Input_maps}
Figure \ref{Fig:CutSky_Info_Input_maps}a shows the CMB simulation A of 
the $\Lambda$CDM concordance model with a pixel resolution of 
$N_{\hbox{\scriptsize side}} = 512$ and an additional smoothing of $10^\circ$. 
In figure \ref{Fig:CutSky_Info_Input_maps}b a second map  
(CMB simulation B) using the 
same cosmological parameters is displayed.
The map in figure \ref{Fig:CutSky_Info_Input_maps}c results from the models 
used for figure \ref{Fig:CutSky_Info_Input_maps}a and 
\ref{Fig:CutSky_Info_Input_maps}b
where the pixels outside the KQ85 mask are taken from simulation A
and inside from simulation B
at the resolution $N_{\hbox{\scriptsize side}} = 512$.
Thereafter a Gaussian smoothing of $10^\circ$ is applied.
}
\end{figure}

\begin{figure}
\begin{center}
{
\begin{minipage}{10cm}
\hspace*{-20pt}\includegraphics[width=9.0cm]{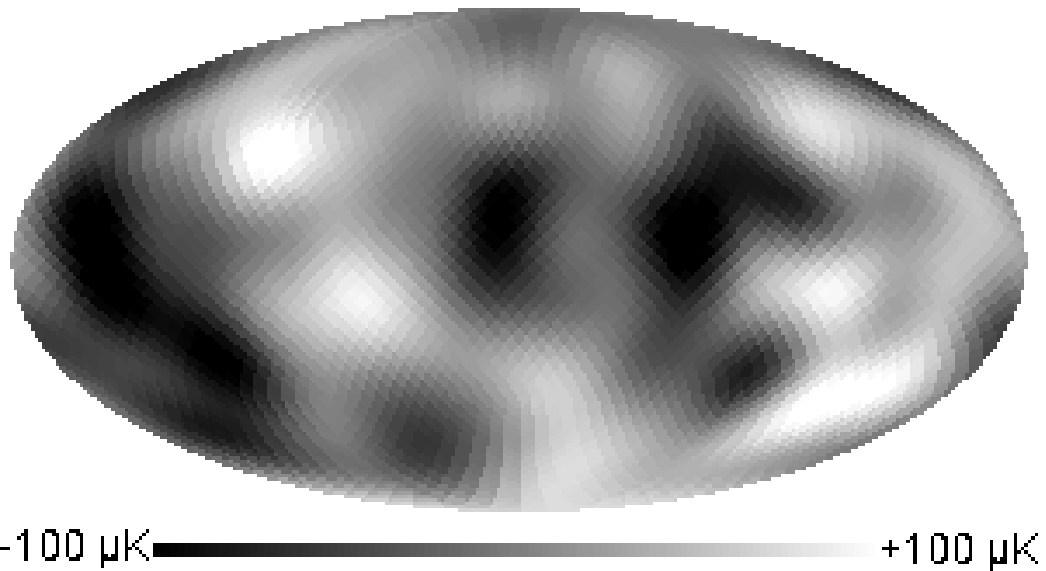}
\end{minipage}
\put(-295,75){(a) $\Lambda$CDM simulation A}
}
\vspace*{10pt}
{
\begin{minipage}{10cm}
\hspace*{-20pt}\includegraphics[width=9.0cm]{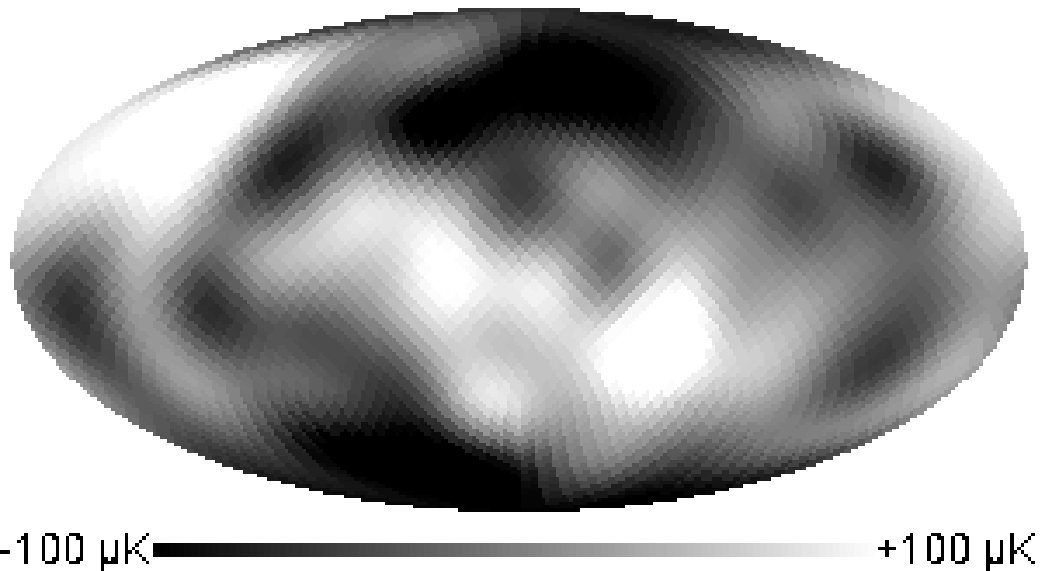}
\end{minipage}
\put(-295,75){(b) $\Lambda$CDM  simulation B}
}
\vspace*{10pt}
{
\begin{minipage}{10cm}
\hspace*{-20pt}\includegraphics[width=9.0cm]{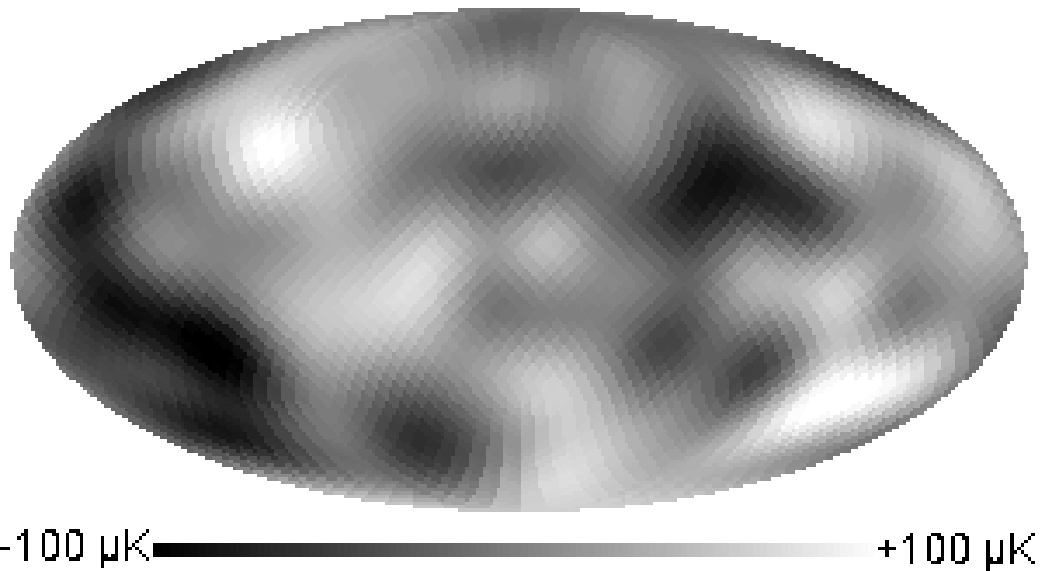}
\end{minipage}
\put(-295,75){(c) $\Lambda$CDM simulation A outside and B inside KQ85 mask}
}
\end{center}
\caption{\label{Fig:CutSky_Info_Extraction_lmax_10}
The three reconstructed maps are displayed.
The reconstructions are carried out for $l_{\hbox{\scriptsize max}}=10$.
These maps result from the maps in figure \ref{Fig:CutSky_Info_Input_maps}a,
b and c by downgrading these maps to a pixel resolution of 
$N_{\hbox{\scriptsize side}} =16$
and thereafter applying the reconstruction algorithm (\ref{Eq:ar_by_A})
to the data outside the mask shown in figure
\ref{Fig:KQ85_masks_nside_512_and_nside_16}b.
Note that panel (a) and (c) should show the same sky map
if the reconstruction algorithm would not use information
from the masked region.
However, there are significant difference especially near the Galactic plane.
}
\end{figure}

In the next sections we will compare results of 
CMB simulations of the $\Lambda$CDM concordance model
with those of the ILC map.
For this reason all maps of the $\Lambda$CDM concordance model are produced
in a FWHM resolution of $1^\circ$ and $N_{\hbox{\scriptsize side}} =512$.
But we also investigate these maps and the ILC map 
after an additional smoothing of e.\,g. $10^\circ$. 
In the following only the additional smoothing width is specified.

Let us now return to the information transfer caused by the
smoothing procedure and/or by carrying out the downgrade.
The following simple numerical experiment 
presented in figures \ref{Fig:CutSky_Info_Input_maps} and
\ref{Fig:CutSky_Info_Extraction_lmax_10} reveals the information transfer.
Figure \ref{Fig:CutSky_Info_Input_maps}a displays
the CMB simulation A of the $\Lambda$CDM concordance model
after a smoothing of $10^\circ$.
This map is downgraded from a pixel resolution of 
$N_{\hbox{\scriptsize side}} =512$ to $N_{\hbox{\scriptsize side}} =16$.
Then the KQ85 mask in the pixel resolution $N_{\hbox{\scriptsize side}} =16$,  
shown in figure \ref{Fig:KQ85_masks_nside_512_and_nside_16}b,
is applied.
For $l_{\hbox{\scriptsize max}}=10$
the reconstruction algorithm (\ref{Eq:ar_by_A})
is used to obtain the reconstructed map
shown in figure \ref{Fig:CutSky_Info_Extraction_lmax_10}a.
Both the original and the reconstructed map agree within the mask
reasonably well.
The same procedure is repeated for a second simulation B
where figure \ref{Fig:CutSky_Info_Input_maps}b shows the simulation
and figure \ref{Fig:CutSky_Info_Extraction_lmax_10}b the reconstruction.
In the next step the pixels of simulation A are replaced
at a pixel resolution of $N_{\hbox{\scriptsize side}} =512$ within the KQ85
mask, shown in figure \ref{Fig:KQ85_masks_nside_512_and_nside_16}a, 
by those of simulation B shown in figure \ref{Fig:CutSky_Info_Input_maps}b.
After this replacement the smoothing of $10^\circ$ is applied.
The last step transfers now the ``wrong'' information to the pixels
outside the mask.
This smoothed map is shown in figure \ref{Fig:CutSky_Info_Input_maps}c.
Downgrading this map to $N_{\hbox{\scriptsize side}} =16$ provides
the data outside the mask which are used for the reconstruction. 
If the reconstruction would not use the information within the mask,
the reconstructed map of figure \ref{Fig:CutSky_Info_Extraction_lmax_10}a
should reappear.
However, as revealed in figure \ref{Fig:CutSky_Info_Extraction_lmax_10}c,
the reconstruction algorithm generates within the mask the main structures
of simulation B, which is displayed in
figure \ref{Fig:CutSky_Info_Input_maps}b.
This clearly demonstrates the information transfer,
so that one has to be careful in testing the reconstruction algorithm.
This leads to the question
whether the reconstruction can be carried out using only unsmoothed maps
where no information about pixels within the mask is encoded outside.
However, then stability difficulties arise as shown below.

\begin{figure}
\begin{center}
\begin{minipage}{11cm}
\hspace*{-20pt}\includegraphics[width=9.0cm]{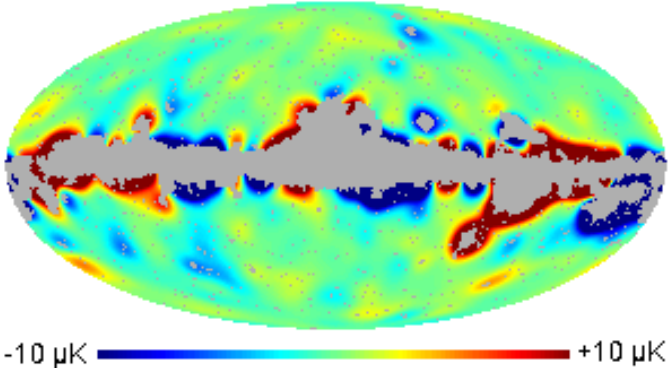}
\end{minipage}
\end{center}
\caption{\label{Fig:wmap_ilc_7yr_v4_leak_600.0arcmin}
The leak of the ILC temperatures from regions within the
mask to those outside due to the smoothing process.
}
\end{figure}

Another way to demonstrate the flow of information from regions within the
mask to those outside the mask, is the following.
At first the pixels of the ILC map outside the KQ85 mask are set to zero.
Then a subsequent $10^\circ$ smoothing shows
how much information about the ILC pixel values inside the mask
leaks to those regions outside.
This leak of information is shown in figure 
\ref{Fig:wmap_ilc_7yr_v4_leak_600.0arcmin}.
It is obvious that the main structures of the ILC map within the mask
appear close to the boundary outside the mask.

The paper is organised as follows. 
In section 2 the reconstruction errors of the CMB temperatures
are evaluated outside and inside the masks using different smoothings and 
different resolutions.
The reconstruction method using the covariance matrix is
compared with the direct inversion method.
In section 3 the influence of the reconstruction method
onto the 2-point correlation function $C(\vartheta)$ of the CMB
is investigated where the focus is on large scales.
An integrated measure of $C(\vartheta)$ serves in section 4
as a further tool to demonstrate the drawbacks of the reconstruction methods.
Finally, in section 5 we summarise our results.

\section{Stability of the reconstruction}

In the last section the quality of the reconstruction method is
tested only visually.
In order to quantify the accuracy of the method,
we start with a temperature map $\delta T_{\hbox{\scriptsize true}}(i)$
which is considered as containing the true temperature information
also inside the mask. 
Here $i$ denotes the index of a pixel within the map.
The reconstruction algorithm (\ref{Eq:ar_by_A}) gets as input only the
temperatures of those pixels lying outside the given mask and returns
a map $\delta T_{\hbox{\scriptsize rec}}(i,l_{\hbox{\scriptsize max}})$
for all pixels.
The resulting map depends on the multipole $l_{\hbox{\scriptsize max}}$
up to which the reconstruction is carried out.
The reconstruction accuracy can then be quantified by
\begin{equation}
\label{Eq:Ortsraum_Test}
\Sigma_{\hbox{\scriptsize rec}}(l_{\hbox{\scriptsize max}}) \, = \,
\sqrt{\frac 1{N_m} {\sum_{i}} '
\left( \delta T_{\hbox{\scriptsize rec}}(i,l_{\hbox{\scriptsize max}}) -
\delta T_{\hbox{\scriptsize true}}(i,l_{\hbox{\scriptsize max}}) \right)^2}
\hspace{5pt} ,
\end{equation}
where the prime means that the sum is restricted to the $N_m$
pixels within the mask.
This gives the average error within the mask.
The magnitude of $\Sigma_{\hbox{\scriptsize rec}}(l_{\hbox{\scriptsize max}})$
depends on the normalisation of the map
$\delta T_{\hbox{\scriptsize true}}(i,l_{\hbox{\scriptsize max}})$,
which is obtained from the original map
$\delta T_{\hbox{\scriptsize true}}(i)$ by taking into account only the
modes up to $l_{\hbox{\scriptsize max}}$.
Note that the expansion
$\delta T_{\hbox{\scriptsize true}}(i,l_{\hbox{\scriptsize max}})$
is only possible for full sky maps since a spherical expansion
is necessary.
To normalise $\Sigma_{\hbox{\scriptsize rec}}(l_{\hbox{\scriptsize max}})$
to a unit fluctuation,
the reconstruction accuracy should be measured with respect to
the mean temperature fluctuation
\begin{equation}
\label{Eq:Ortsraum_Normierung}
\sigma_{\hbox{\scriptsize true}}(l_{\hbox{\scriptsize max}}) \; = \;
\sqrt{\frac 1{N_o} {\sum_{i}} ''
\delta T_{\hbox{\scriptsize true}}^2(i,l_{\hbox{\scriptsize max}})}
\hspace{10pt} ,
\end{equation}
where the sum with the two primes is restricted to the $N_o$
pixels outside the mask.
The normalised total reconstruction error is then given by the quotient
\begin{equation}
\label{Eq:Quality}
Q(l_{\hbox{\scriptsize max}}) \; = \;
\frac{\Sigma_{\hbox{\scriptsize rec}}(l_{\hbox{\scriptsize max}})}
{\sigma_{\hbox{\scriptsize true}}(l_{\hbox{\scriptsize max}})}
\hspace{10pt} .
\end{equation}
A value of $Q=1$ means that the error is as large as a typical
temperature value,
i.\,e.\ the reconstruction is useless.

\begin{figure}
\vspace*{-30pt}
\begin{center}
{
\begin{minipage}{11cm}
\hspace*{-20pt}\includegraphics[width=10.0cm]{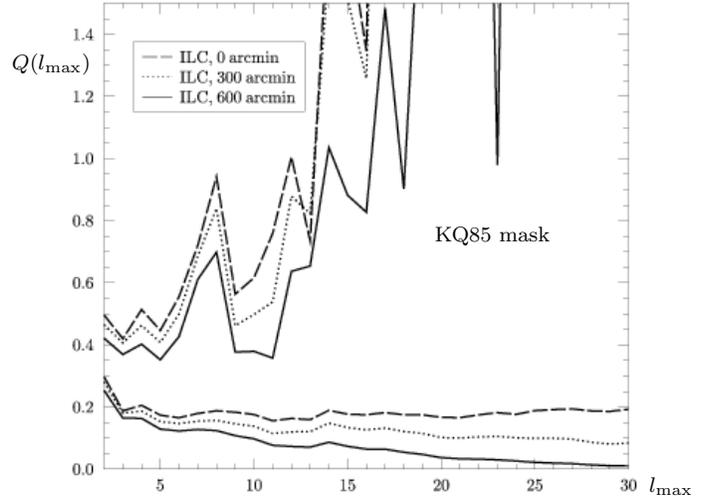}
\end{minipage}
\put(-325,65){$Q(l_{\hbox{\scriptsize max}})$}
\put(-84,-95){$l_{\hbox{\scriptsize max}}$}
\put(-165,0){KQ85 mask}
}
\end{center}
\vspace*{-20pt}
\caption{\label{Fig:Q_ilc_KQ85}
The three upper curves show the normalised  total error
$Q(l_{\hbox{\scriptsize max}})$
of the reconstruction applied to the ILC (7yr) map using the KQ85 (7yr) mask
in dependence on the multipole $l_{\hbox{\scriptsize max}}$.
In addition to the original ILC map, the algorithm is applied to
smoothed ILC maps with a smoothing of 300 and 600 arcmin at the resolution
$N_{\hbox{\scriptsize side}} = 512$.
The reconstruction is carried out for the mask threshold
$x_{\hbox{\scriptsize th}}=0.5$ 
using the resolution $N_{\hbox{\scriptsize side}} = 16$.
The three lower curves show the  total error
$Q(l_{\hbox{\scriptsize max}})$
evaluated for the pixels outside the mask.
}
\end{figure}

\begin{figure}
\begin{center}
{
\begin{minipage}{11cm}
\hspace*{-20pt}\includegraphics[width=9.0cm]{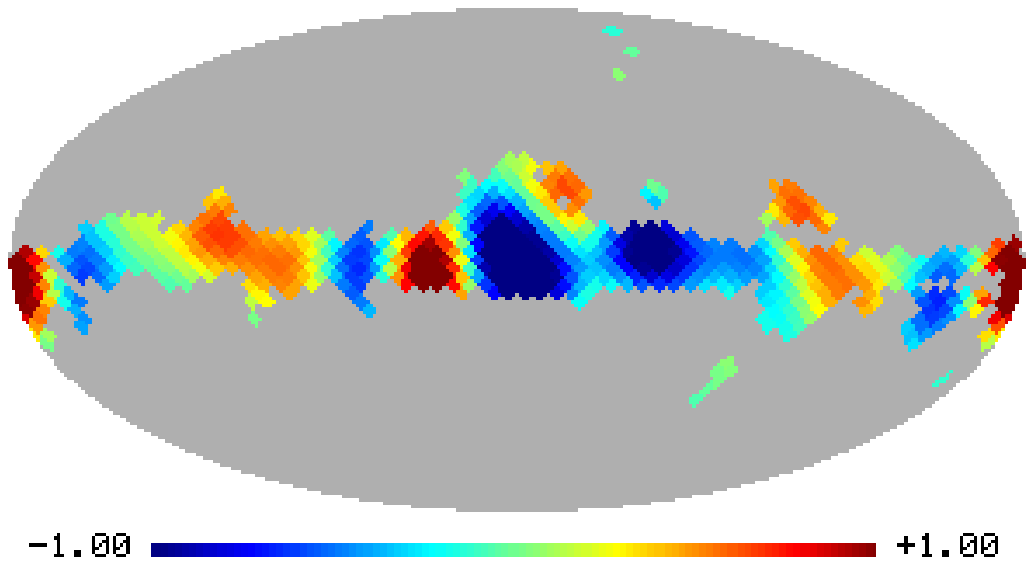}
\end{minipage}
\put(-330,70){(a) inside KQ85 mask}
}
\vspace{10pt}
{
\begin{minipage}{11cm}
\hspace*{-20pt}\includegraphics[width=9.0cm]{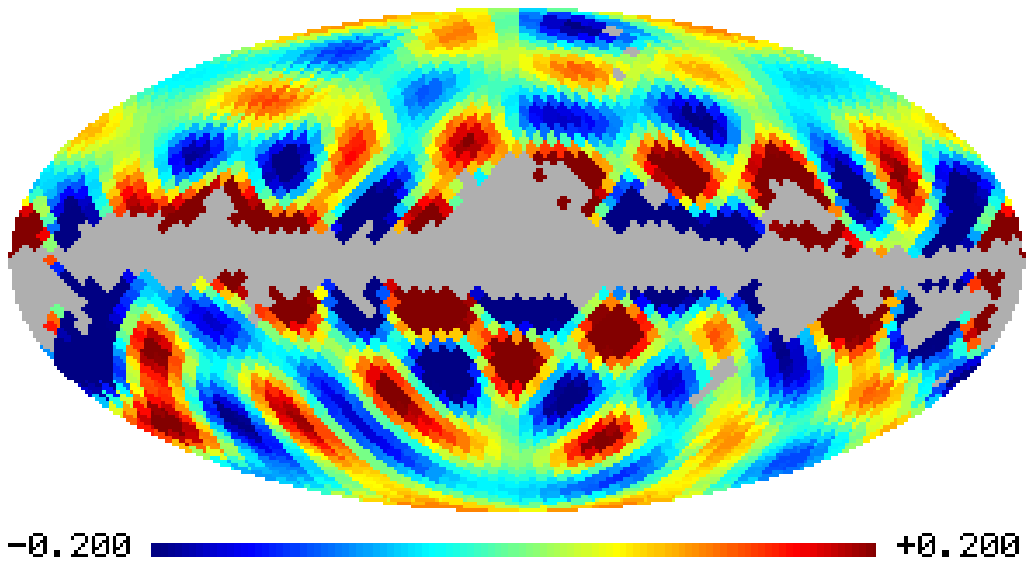}
\end{minipage}
\put(-330,70){(b) outside KQ85 mask}
}
\end{center}
\caption{\label{Fig:q_ilc_KQ85_lmax_10}
The upper panel shows the local reconstruction error (\ref{Eq:local_Quality})
of the ILC  (7yr) map within the KQ85 mask corresponding to
the  total reconstruction error $Q(10)=0.62$ of the ILC map as can be read off
from figure \ref{Fig:Q_ilc_KQ85} at $l_{\hbox{\scriptsize max}}=10$.
The colour bar is truncated such that
$q$-values less than -1.0 are pictured as dark blue 
and larger than 1.0 as dark red.
In the lower panel the same quantity is pictured outside the mask. 
Here typical $q$-values lie between -0.2 and +0.2
as can be read off in figure \ref{Fig:Q_ilc_KQ85}. 
For this reason the maximum and minimum values are truncated at 
0.2 and -0.2, respectively.
}
\end{figure}

We apply the normalised accuracy measure (\ref{Eq:Quality})
to the ILC (7yr) map and to simulations of
the $\Lambda$CDM concordance model where the cosmological parameters
of the WMAP team are used 
(Table 8, column ``WMAP+BAO+$H_0$'' in \cite{Jarosik_et_al_2010}).
These maps with a resolution of $N_{\hbox{\scriptsize side}} = 512$
are smoothed to a Gaussian width $\theta = 1^{\circ}$, 
and in the following only a smoothing in addition to this
$\theta = 1^{\circ}$ smoothing is given. 
After the smoothing the maps are downgraded to lower resolutions
of $N_{\hbox{\scriptsize side}} = 16$ or 32.
Then the downgraded masks are applied in this lower resolution,
which gives the maps which serve as the input for the
reconstruction algorithm.

In figure \ref{Fig:Q_ilc_KQ85} the three upper curves reveal
the  total reconstruction error (\ref{Eq:Quality})
for ILC (7yr) map using the KQ85 (7yr) mask for different smoothings.
As outlined above the algorithm depends
on the multipole $l_{\hbox{\scriptsize max}}$ up to which
the multipoles $l = 0,\dots,l_{\hbox{\scriptsize max}}$
are to be reconstructed.
The accuracy measure (\ref{Eq:Quality}) refers to the
maps expanded only up to $l=l_{\hbox{\scriptsize max}}$.
Even for very low values of $l_{\hbox{\scriptsize max}} \lesssim 10$
the  total error $Q(l_{\hbox{\scriptsize max}})$ is above 0.4,
i.\,e.\ the temperatures reconstructed within the mask
differ from the true ones by one half of the mean temperature
fluctuations.
This demonstrates the accuracy problems of the algorithm.
It is seen that the error decreases with increasing Gaussian smoothing
because the information leak is getting stronger with increasing smoothing.
The three lower curves shown in figure \ref{Fig:Q_ilc_KQ85}
are the  total errors outside the mask.
These are computed using equation (\ref{Eq:Ortsraum_Test})
but with the modification that the primed sum runs only over those
pixels $i$ outside the mask.
It is seen that with increasing value of 
$l_{\hbox{\scriptsize max}}$  the accuracy gets better
although the error is surprisingly large for small
reconstruction multipoles $l_{\hbox{\scriptsize max}}$.
With additional smoothing the  total reconstruction error of the ILC map
has the tendency to fall, but without additional smoothing the  total error
is nearly constant about 0.2.

Up to now only the global accuracy measure $Q$ is discussed.
But in order to find the regions which are most prone to reconstruction errors,
we now introduce the reconstruction accuracy of single pixels
which is obtained by taking the difference between
$\delta T_{\hbox{\scriptsize rec}}(i,l_{\hbox{\scriptsize max}})$ and
$\delta T_{\hbox{\scriptsize true}}(i,l_{\hbox{\scriptsize max}})$.
Normalising this difference with
$\sigma_{\hbox{\scriptsize true}}(l_{\hbox{\scriptsize max}})$
results in the local reconstruction error
\begin{equation}
\label{Eq:local_Quality}
q(i,l_{\hbox{\scriptsize max}}) \; = \;
\frac{ \delta T_{\hbox{\scriptsize rec}}(i,l_{\hbox{\scriptsize max}}) -
\delta T_{\hbox{\scriptsize true}}(i,l_{\hbox{\scriptsize max}})}
{\sigma_{\hbox{\scriptsize true}}(l_{\hbox{\scriptsize max}})}
\hspace{10pt}.
\end{equation}

To expose the regions with the largest errors,
the local reconstruction error (\ref{Eq:local_Quality})
of the ILC map (without additional smoothing) 
is pictured in figure \ref{Fig:q_ilc_KQ85_lmax_10}
for $l_{\hbox{\scriptsize max}}=10$. 
The errors within the mask are shown in figure \ref{Fig:q_ilc_KQ85_lmax_10}a. 
In this figure dark areas display errors with $|q|\ge 1$. 
Such values of $q$ occur near the galactic centre 
where the mask is very extended.
But errors larger than 1 also occur antipodal to the galactic centre.
The errors outside the mask are shown in figure \ref{Fig:q_ilc_KQ85_lmax_10}b.
Based on figure \ref{Fig:Q_ilc_KQ85} one would expect that
a typical value of the local reconstruction error (\ref{Eq:local_Quality}) 
is about $\pm 0.2$.
The colour band is truncated at those values of $|q|$
such that values of $|q|\ge 0.2$ are pictured as dark blue or dark red.
It is seen that large errors outside the mask
appear near the masked region.
A further observation is that with increasing $l_{\hbox{\scriptsize max}}$  
the most significant errors approach the masked region.
Then the values of the maximum errors increase,
although the areas with $|q|\ge 0.2$ are getting smaller.

\begin{figure}
\begin{center}
\vspace*{-30pt}
{
\begin{minipage}{11cm}
\hspace*{-20pt}\includegraphics[width=10.0cm]{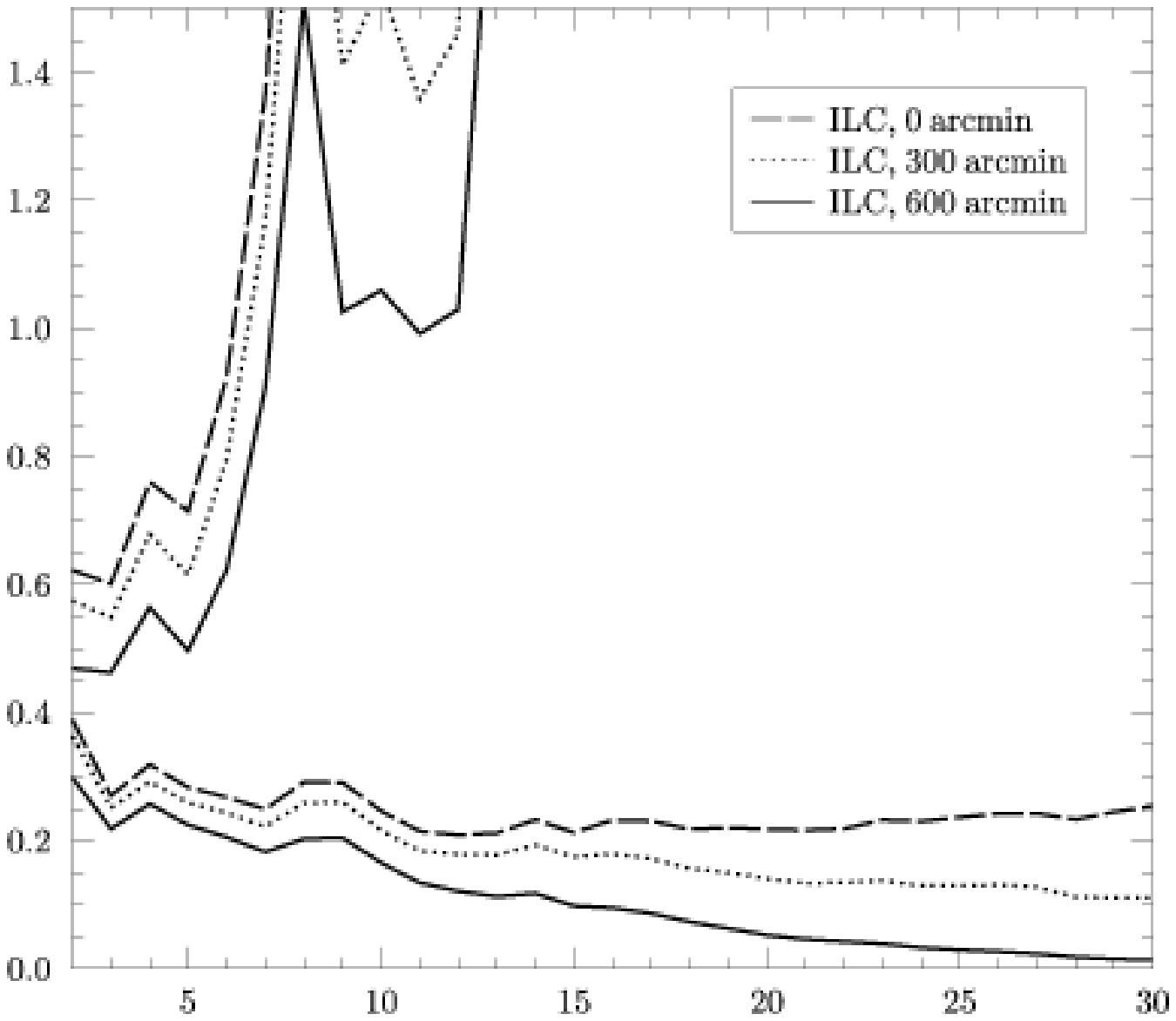}
\end{minipage}
\put(-325,65){$Q(l_{\hbox{\scriptsize max}})$}
\put(-84,-95){$l_{\hbox{\scriptsize max}}$}
\put(-165,0){KQ75 mask}
}
\end{center}
\vspace*{-20pt}
\caption{\label{Fig:Q_ilc_KQ75}
The same analysis for the normalised total error
$Q(l_{\hbox{\scriptsize max}})$ as in figure \ref{Fig:Q_ilc_KQ85}
is presented,
but now the larger KQ75 (7yr) mask is used instead of the KQ85 (7yr) mask.
}
\end{figure}

\begin{figure}
\begin{center}
{
\begin{minipage}{11cm}
\hspace*{-20pt}\includegraphics[width=9.0cm]{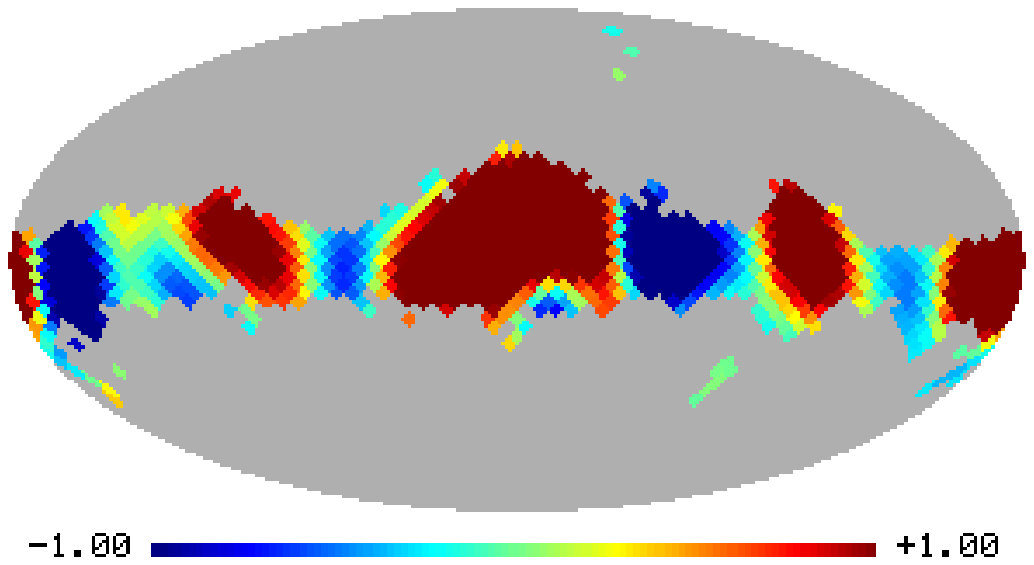}
\end{minipage}
\put(-330,70){(a) inside KQ75 mask}
}
\vspace{10pt}
{
\begin{minipage}{11cm}
\hspace*{-20pt}\includegraphics[width=9.0cm]{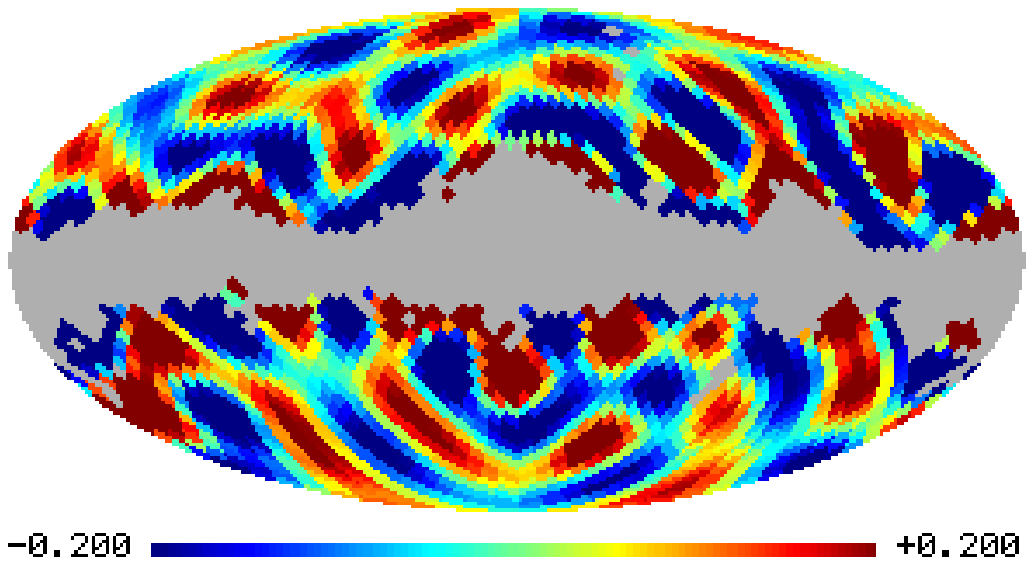}
\end{minipage}
\put(-330,70){(b) outside KQ75 mask}
}
\end{center}
\caption{\label{Fig:q_ilc_KQ75_lmax_10}
The upper panel depicts the local reconstruction error (\ref{Eq:local_Quality})
of the ILC (7yr) map within the KQ75 mask corresponding to
the  total reconstruction error $Q(10)=1.86$
as figure \ref{Fig:Q_ilc_KQ75} reveals at $l_{\hbox{\scriptsize max}}=10$.
In the lower panel the same quantity is pictured outside the mask. 
The colour bands are truncated in the same way as in figure
\ref{Fig:q_ilc_KQ85_lmax_10}, i.\,e.\ at $\pm 1.0$ and $\pm 0.2$
in the upper and lower panel, respectively.
}
\end{figure}

Since the reconstruction error depends on the size of the mask,
it is interesting to consider the larger KQ75 (7yr) mask
which again is applied to the ILC (7yr) map.
In figure \ref{Fig:Q_ilc_KQ75} the three upper curves display
the corresponding total reconstruction errors (\ref{Eq:Quality})
within the mask.
Even for very low values of $l_{\hbox{\scriptsize max}} \lesssim 5$
the  total error $Q(l_{\hbox{\scriptsize max}})$ is above 0.5,
i.\,e.\ the temperatures reconstructed within the mask differ
from the original ones by more than one half of the mean temperature
fluctuation.
For $l_{\hbox{\scriptsize max}} \ge 7$ 
the  total error $Q(l_{\hbox{\scriptsize max}})$ is even above 1.0.
It is seen that the error decreases with increasing Gaussian smoothing
because of the information leak, 
but nevertheless the  total error $Q(l_{\hbox{\scriptsize max}})$ is large
for a smoothing with 600 arcmin.
The three lower curves shown in figure \ref{Fig:Q_ilc_KQ75}
are the total errors outside the mask.
It is seen that with increasing value of 
$l_{\hbox{\scriptsize max}}$  the accuracy improves
although the error is surprisingly large for small
reconstruction multipoles $l_{\hbox{\scriptsize max}}$.
With additional smoothing the total reconstruction error of the ILC map
has the tendency to fall, but without additional smoothing the error
is nearly constant at about 0.25.

To reveal the locations of the most significant errors 
in the case of the KQ75 (7yr) mask,
the local reconstruction error of the ILC map (without additional smoothing) 
is pictured in figure \ref{Fig:q_ilc_KQ75_lmax_10}
for $l_{\hbox{\scriptsize max}}=10$.
The errors within the mask are shown in figure \ref{Fig:q_ilc_KQ75_lmax_10}a.
The dark areas display errors with $|q|\ge 1$
which are much more widespread than for the KQ85 (7yr) mask. 
The errors outside the mask are shown in figure \ref{Fig:q_ilc_KQ75_lmax_10}b.
To locate large errors we have pictured values of $|q|\ge 0.2$
as dark blue or dark red.
The errors outside the mask are only slightly larger than
in the case of the KQ85 (7yr) mask.

Compared to the KQ85 (7yr) mask, the accuracy problems of the algorithm
applied to the larger KQ75 (7yr) mask are much stronger,
and we conclude that this mask is too large in order
to allow a reconstruction for $l_{\hbox{\scriptsize max}} \ge 7$,
and even the results for $l_{\hbox{\scriptsize max}} < 7$ 
should be considered critical.
For this reason we concentrate us in the following
on reconstructions using the smaller KQ85 (7yr) mask.

\begin{figure}
\begin{center}
\vspace*{-15pt}
{
\begin{minipage}{11cm}
\includegraphics[width=8.5cm]{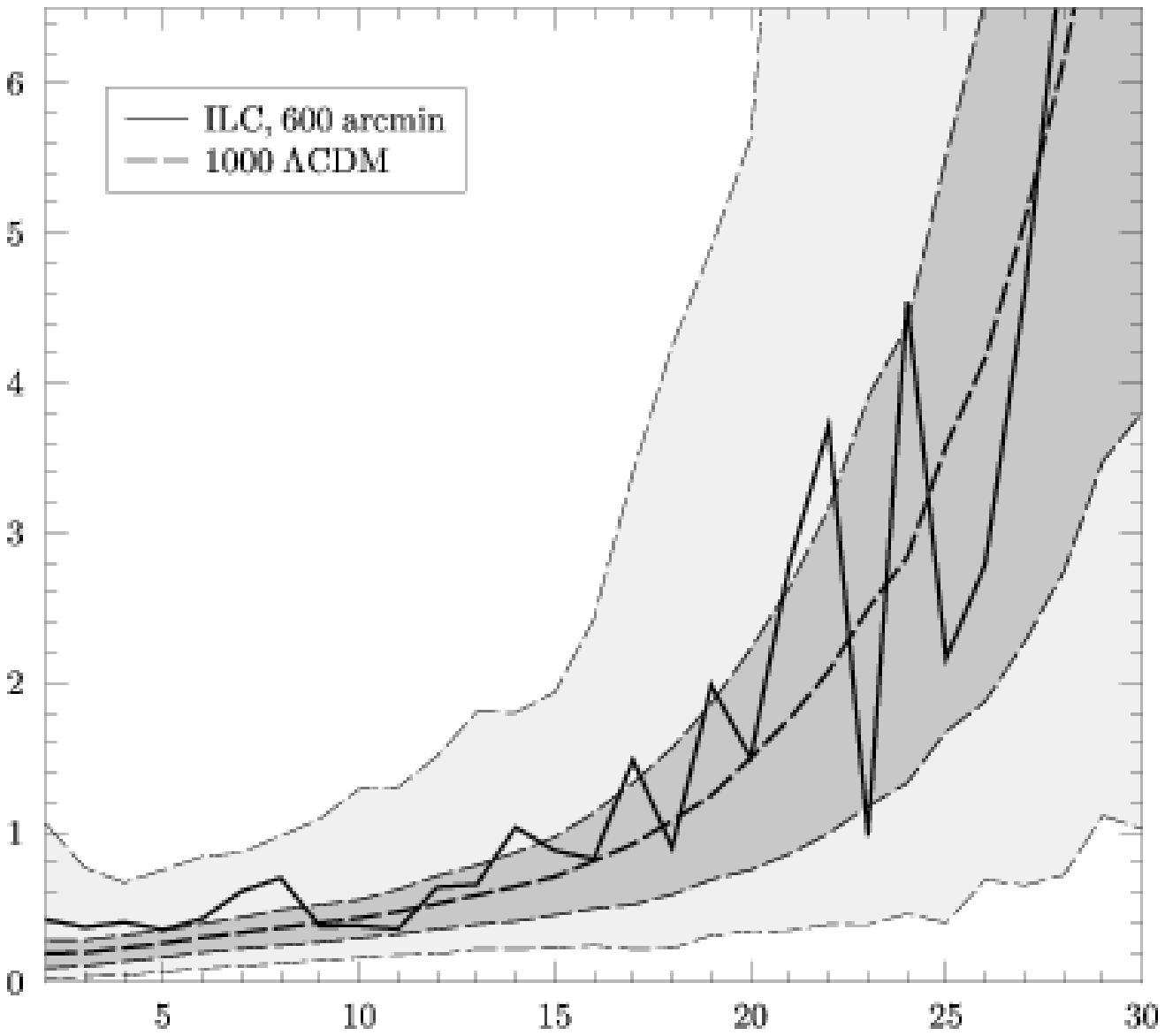}
\end{minipage}
\put(-325,65){$Q(l_{\hbox{\scriptsize max}})$}
\put(-84,-95){$l_{\hbox{\scriptsize max}}$}
\put(-275,45){(a) inside KQ85 mask}
}
\vspace*{-10pt}
{
\begin{minipage}{11cm}
\includegraphics[width=8.5cm]{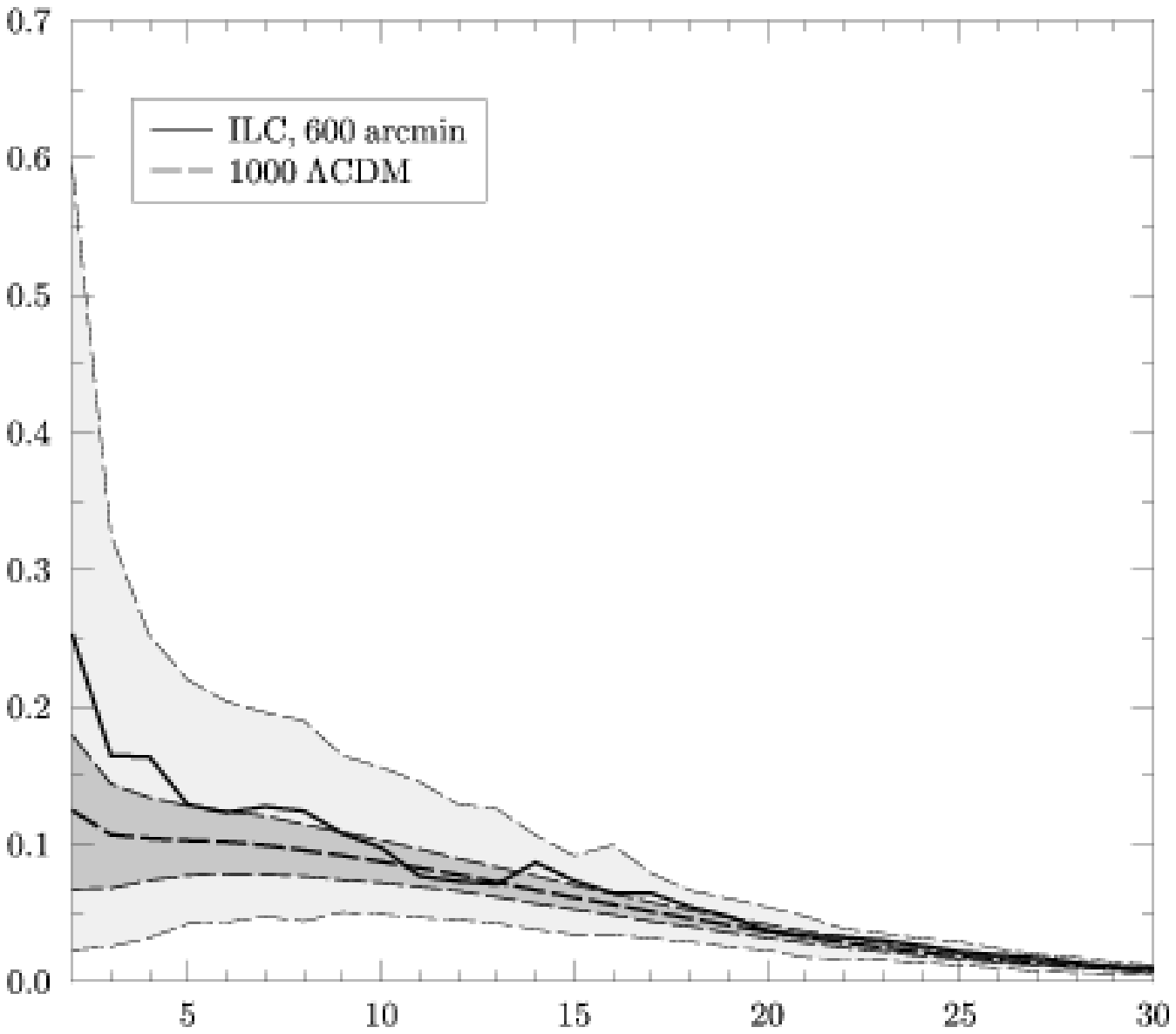}
\end{minipage}
\put(-325,75){$Q(l_{\hbox{\scriptsize max}})$}
\put(-84,-95){$l_{\hbox{\scriptsize max}}$}
\put(-275,45){(b) outside KQ85 mask}
}
\end{center}
\vspace*{-10pt}
\caption{\label{Fig:Q_ilc_und_1000lcdm_KQ85_FWHM_600arcmin}
In the upper panel the solid curve shows the normalised total error
$Q(l_{\hbox{\scriptsize max}})$
of the reconstruction inside the KQ85 (7yr) mask 
applied to the ILC (7yr) map smoothed with 600 arcmin
in dependence on the multipole $l_{\hbox{\scriptsize max}}$.
The reconstruction is carried out for the
mask threshold $x_{\hbox{\scriptsize th}}=0.5$ 
using the resolution $N_{\hbox{\scriptsize side}} = 16$.
The mean value of the total reconstruction error 
calculated from 1000 CMB simulations is displayed as a dashed curve.
The range containing 66.6 \% of these models is pictured as dark grey band. 
The distribution of all errors of these 1000 CMB simulations is given as 
a light grey band.
The lower panel shows the corresponding total 
reconstruction errors outside the mask.
}
\end{figure}

\begin{figure}
\begin{center}
\vspace*{-15pt}
{
\begin{minipage}{11cm}
\includegraphics[width=8.5cm]{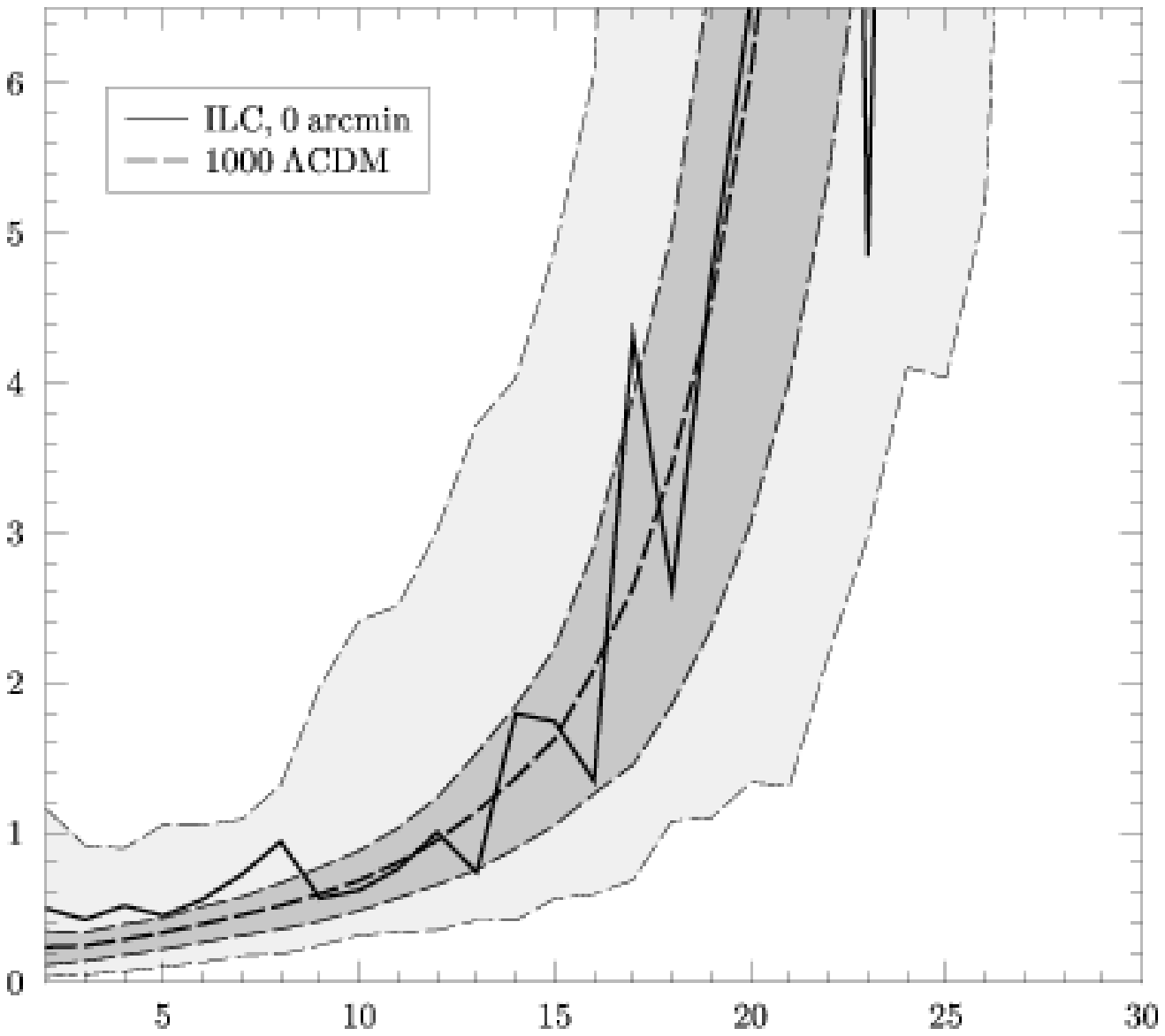}
\end{minipage}
\put(-325,65){$Q(l_{\hbox{\scriptsize max}})$}
\put(-84,-95){$l_{\hbox{\scriptsize max}}$}
\put(-285,45){(a) inside KQ85 mask}
}
\vspace*{-10pt}
{
\begin{minipage}{11cm}
\includegraphics[width=8.5cm]{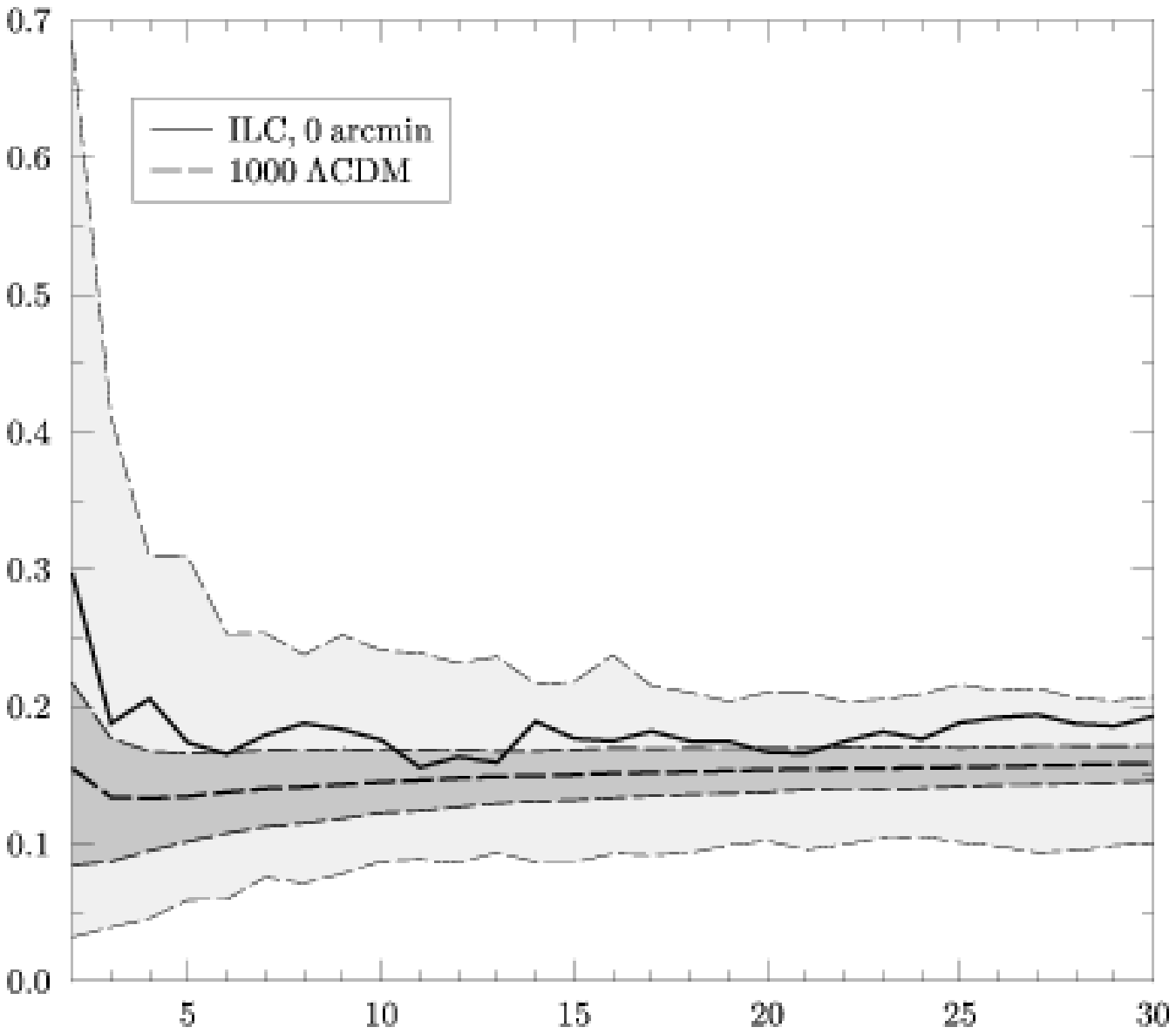}
\end{minipage}
\put(-325,75){$Q(l_{\hbox{\scriptsize max}})$}
\put(-84,-95){$l_{\hbox{\scriptsize max}}$}
\put(-285,45){(b) outside KQ85 mask}
}
\end{center}
\vspace*{-10pt}
\caption{\label{Fig:Q_ilc_und_1000lcdm_KQ85_FWHM_0arcmin}
The same analysis of the reconstruction error $Q(l_{\hbox{\scriptsize max}})$
as in figure \ref{Fig:Q_ilc_und_1000lcdm_KQ85_FWHM_600arcmin}
is shown, but now without smoothing.
}
\end{figure}

Now we compare the total reconstruction error of the ILC map
with those of 1000 CMB realizations of the $\Lambda$CDM concordance model.
In figure \ref{Fig:Q_ilc_und_1000lcdm_KQ85_FWHM_600arcmin}
this comparison is shown with the additional smoothing of 600 arcmin
and in figure \ref{Fig:Q_ilc_und_1000lcdm_KQ85_FWHM_0arcmin}
without smoothing.
A resolution of $N_{\hbox{\scriptsize side}} = 16$ and
a mask threshold $x_{\hbox{\scriptsize th}}=0.5$ is used.
As can be seen in the upper panel of
figure \ref{Fig:Q_ilc_und_1000lcdm_KQ85_FWHM_600arcmin}
the total reconstruction error of the ILC map within the mask is 
located for $l_{\hbox{\scriptsize max}}\le 8$ outside 
the dark grey band containing 66.6\% of 1000 models,
but between the maximum and minimum errors of these models
which domain is depicted as the light grey band.
For $l_{\hbox{\scriptsize max}} > 8$
the total reconstruction error of the ILC map is rather typical.
It should be noted that the computation of the
ILC reconstruction error assumes that the ILC map contains
the true values inside the mask
which, of course, needs not to be the case.
But as the comparison with the light grey band shows,
the total reconstruction error is always smaller than the extreme errors
that occur among the 1000 simulations.
The total reconstruction error from these 1000 models
with an additional smoothing of 600 arcmin reveals
that the reconstruction of the CMB within the mask is impossible
for $l_{\hbox{\scriptsize max}} \gtrsim 16$.
In figure \ref{Fig:Q_ilc_und_1000lcdm_KQ85_FWHM_600arcmin}b 
the total reconstruction error of the ILC map outside the mask is compared
with those of the 1000 $\Lambda$CDM models.
For $l_{\hbox{\scriptsize max}}\le 8$
the total reconstruction error of the ILC map
is again outside the dark grey band containing 66.6\% of the simulations.
Furthermore, with increasing $l_{\hbox{\scriptsize max}}$
the total reconstruction error outside the mask vanishes
if the additional smoothing of 600 arcmin is applied.

Figure \ref{Fig:Q_ilc_und_1000lcdm_KQ85_FWHM_0arcmin} displays
the total reconstruction error without the additional smoothing
of figure \ref{Fig:Q_ilc_und_1000lcdm_KQ85_FWHM_600arcmin}.
The total reconstruction error within the mask
of the ILC map and of the 1000 simulations
is larger without additional smoothing.
With increasing value of $l_{\hbox{\scriptsize max}}$
the reconstruction of the temperature fluctuations within the mask
deteriorates increasingly,
such that the reconstruction is doubtful for
$l_{\hbox{\scriptsize max}} \gtrsim 13$.
The mean values of the total reconstruction error outside the mask of
the 1000 $\Lambda$CDM simulations are now nearly
independent from $l_{\hbox{\scriptsize max}}$ at a value about 0.15.
The width of the distribution of the total reconstruction error decreases
with $l_{\hbox{\scriptsize max}}$.

\begin{figure}
\begin{center}
{
\begin{minipage}{11cm}
\hspace*{-20pt}\includegraphics[width=9.0cm]{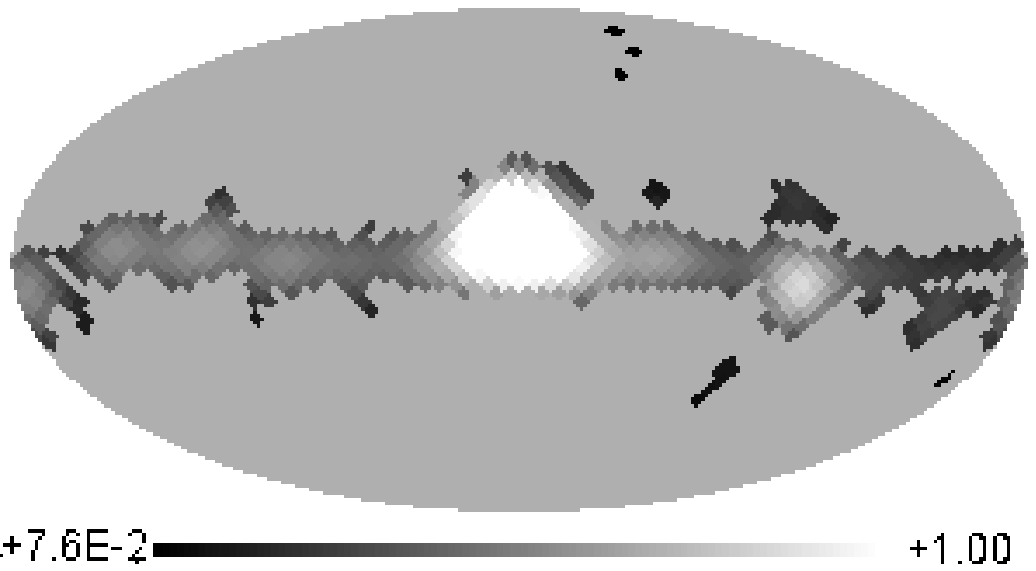}
\end{minipage}
\put(-330,70){(a) inside KQ85 mask}
}
\vspace{10pt}
{
\begin{minipage}{11cm}
\hspace*{-20pt}\includegraphics[width=9.0cm]{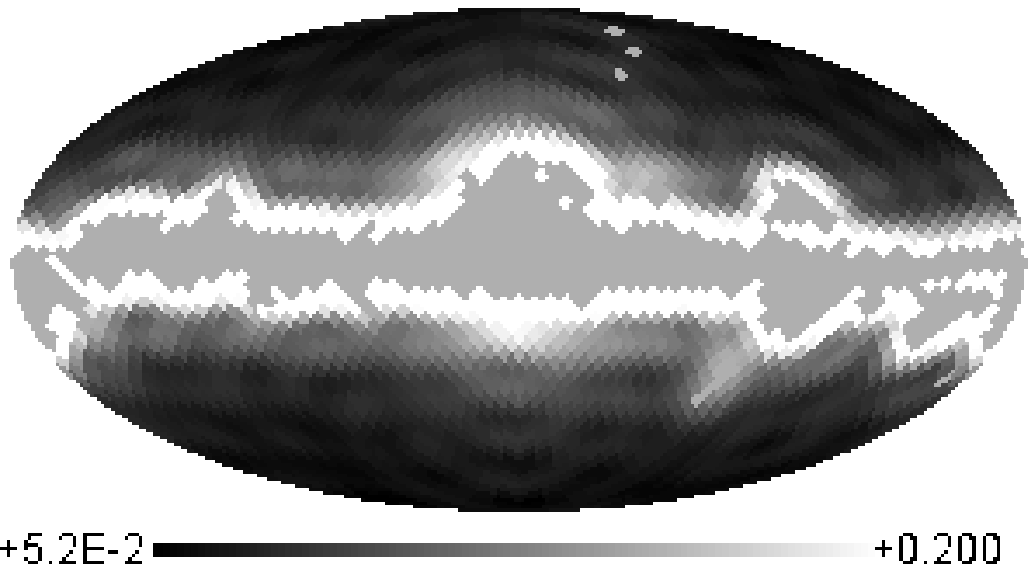}
\end{minipage}
\put(-330,70){(b) outside KQ85 mask}
}
\end{center}
\caption{\label{Fig:q_lcdm_KQ85_lmax_10}
The standard deviation of the local reconstruction error
(\ref{Eq:local_Quality}) is displayed for $l_{\hbox{\scriptsize max}}=10$
calculated from 1000 $\Lambda$CDM maps. 
The upper panel depicts the standard deviation of
$q(i,l_{\hbox{\scriptsize max}})$ within the mask,
where values $q \ge 1$ are pictured white. 
In the centre of the galaxy the standard deviation has values
as large as 2.12.
At the lower panel the same quantity is pictured outside the mask,
but now the maximum value is truncated at 0.2. 
Pixel values larger than 0.4 occur.
} 
\end{figure}

Figures \ref{Fig:q_ilc_KQ85_lmax_10} and \ref{Fig:q_ilc_KQ75_lmax_10}
depict the local reconstruction error $q(i,l_{\hbox{\scriptsize max}})$
for the ILC map, where one has to assume
that the ILC map contains the true pixel values within the mask.
The locations of extreme values of $q(i,l_{\hbox{\scriptsize max}})$
without this assumption are obtained from the 1000 $\Lambda$CDM simulations.
The standard deviation of $q(i,l_{\hbox{\scriptsize max}})$
calculated from these 1000 models (without additional smoothing)
is pictured in figure \ref{Fig:q_lcdm_KQ85_lmax_10}
for $l_{\hbox{\scriptsize max}}=10$.
The upper panel shows the standard deviation within the mask
and the lower one outside the mask.
Within the mask the standard deviation is extremely large near the centre of
the galaxy where a huge area is masked.
In general, one finds that the larger the masked domain is,
the larger is the standard deviation of $q(i,l_{\hbox{\scriptsize max}})$
in the centre of such a domain,
and accordingly, the larger is the expected local reconstruction error 
for an individual map. 
Outside the mask, extreme standard deviations of
$q(i,l_{\hbox{\scriptsize max}})$ occur near the boundary of the mask.
The comparison of the local reconstruction error
in figure \ref{Fig:q_ilc_KQ85_lmax_10} with this standard deviation
shows that antipodal to the centre of the galaxy
the local reconstruction error is with about 4$\sigma$ untypically large.
This could be a hint for residual foreground in the ILC map.

\begin{figure}
\begin{center}
\vspace*{-15pt}
{
\begin{minipage}{11cm}
\includegraphics[width=8.5cm]{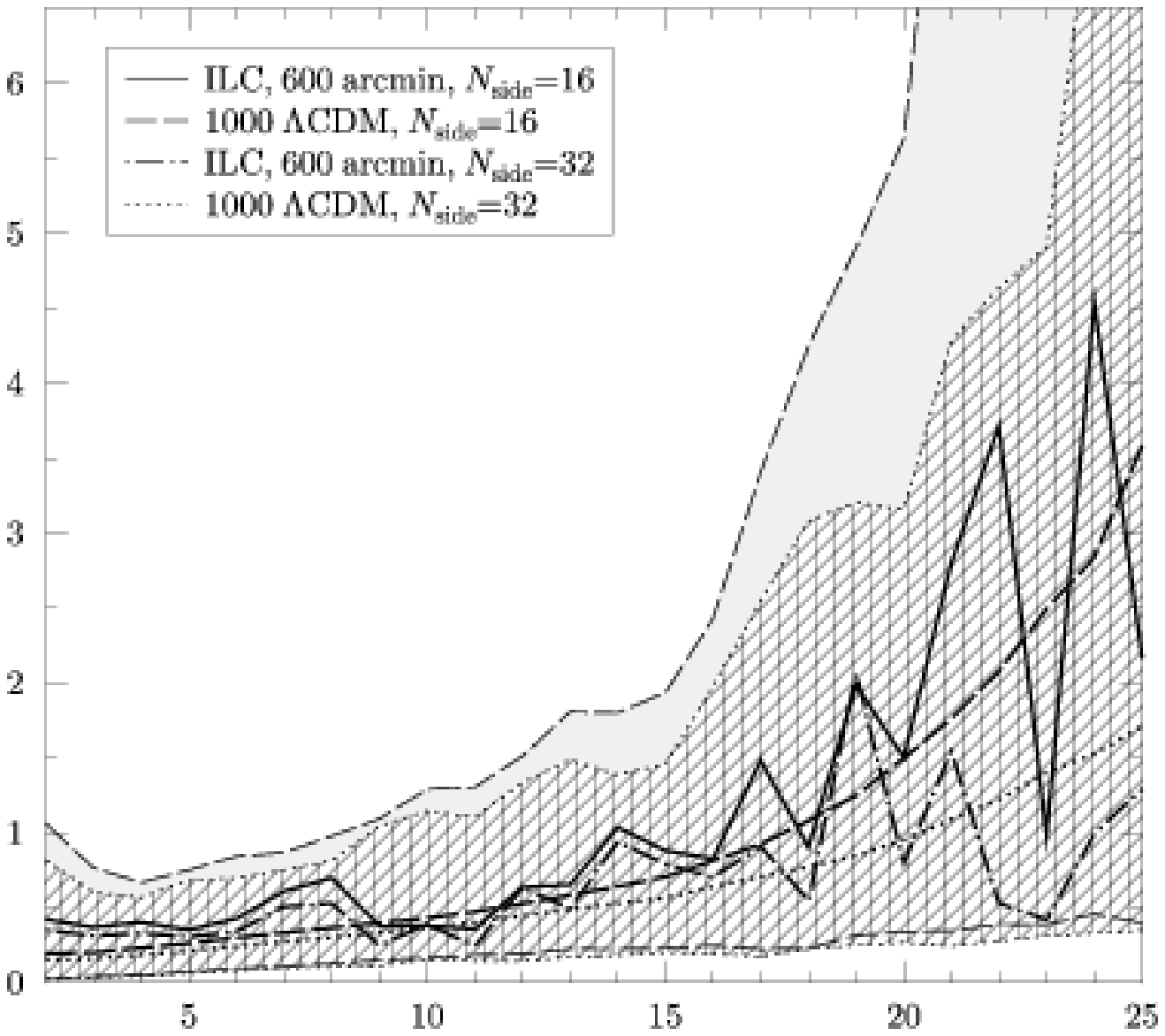}
\end{minipage}
\put(-325,65){$Q(l_{\hbox{\scriptsize max}})$}
\put(-84,-95){$l_{\hbox{\scriptsize max}}$}
\put(-275,38){(a) inside KQ85 mask}
}
\vspace*{-10pt}
{
\begin{minipage}{11cm}
\includegraphics[width=8.5cm]{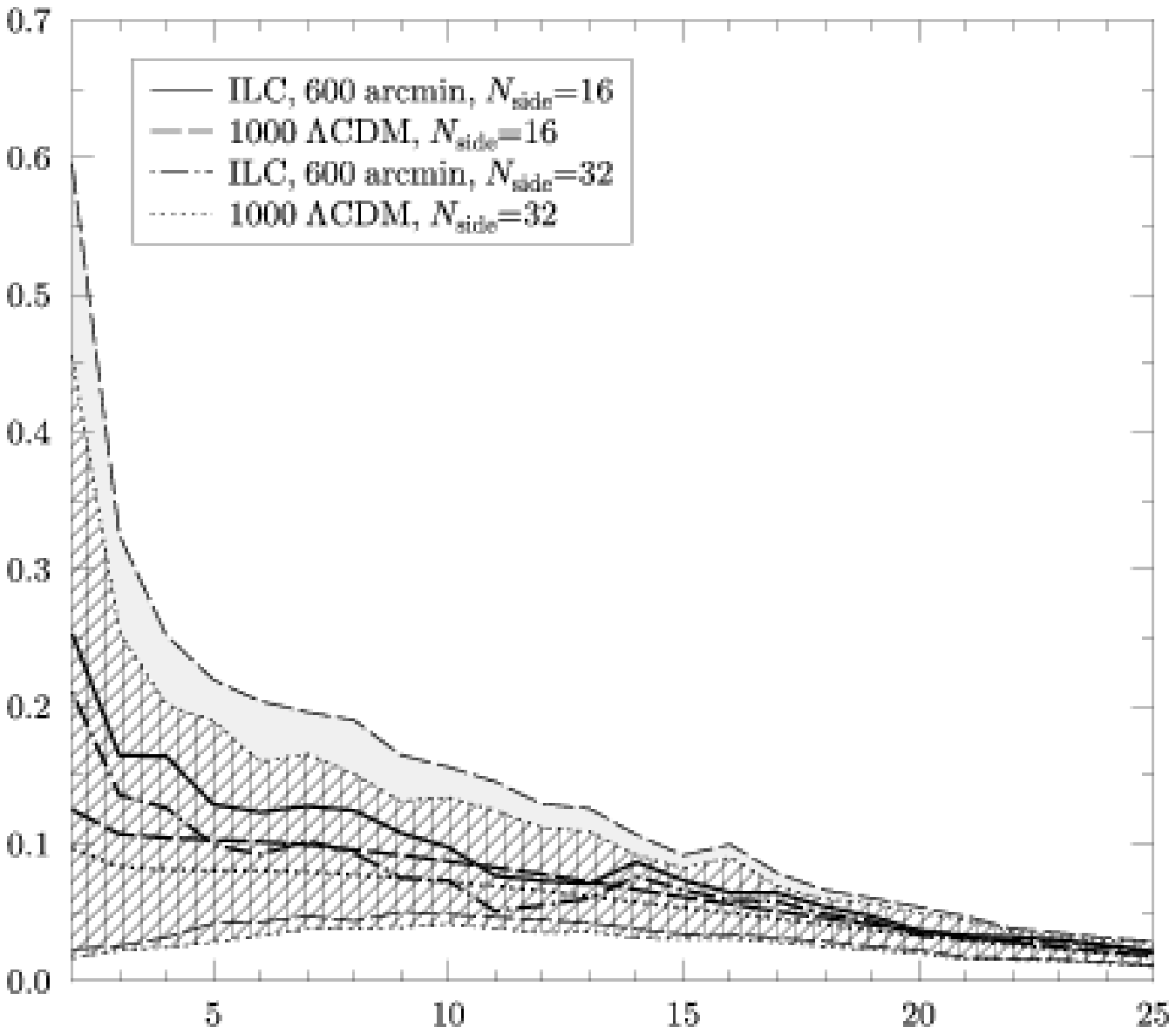}
\end{minipage}
\put(-325,75){$Q(l_{\hbox{\scriptsize max}})$}
\put(-84,-95){$l_{\hbox{\scriptsize max}}$}
\put(-275,38){(b) outside KQ85 mask}
}
\end{center}
\vspace*{-10pt}
\caption{\label{Fig:Q_ilc_und_1000lcdm_KQ85_FWHM_600arcmin_nside_16_und_32}
The normalised total error $Q(l_{\hbox{\scriptsize max}})$
is computed for the two different resolutions of
$N_{\hbox{\scriptsize side}} = 16$ and $N_{\hbox{\scriptsize side}} = 32$.
The total error $Q(l_{\hbox{\scriptsize max}})$ is obtained from
the ILC (7yr) map together with KQ85 (7yr) mask using a threshold
$x_{\hbox{\scriptsize th}}=0.5$.
A smoothing of 600 arcmin is applied.
The total error for the ILC (7yr) map for $N_{\hbox{\scriptsize side}} = 16$ 
is plotted as a solid curve and for $N_{\hbox{\scriptsize side}} = 32$ as
a dash-dotted curve.
This error is compared with the corresponding mean error obtained from
1000 $\Lambda$CDM simulations
which is displayed as a dashed curve and as a dotted line, respectively.  
The distribution of the errors of the 1000 CMB simulations is given as
a light grey band and as a shaded band for
$N_{\hbox{\scriptsize side}} = 16$ and $N_{\hbox{\scriptsize side}} = 32$,
respectively.
The upper and lower panels show the total reconstruction errors
inside and outside the mask.
}
\end{figure}

\begin{figure}
\begin{center}
\vspace*{-15pt}
{
\begin{minipage}{11cm}
\includegraphics[width=8.5cm]{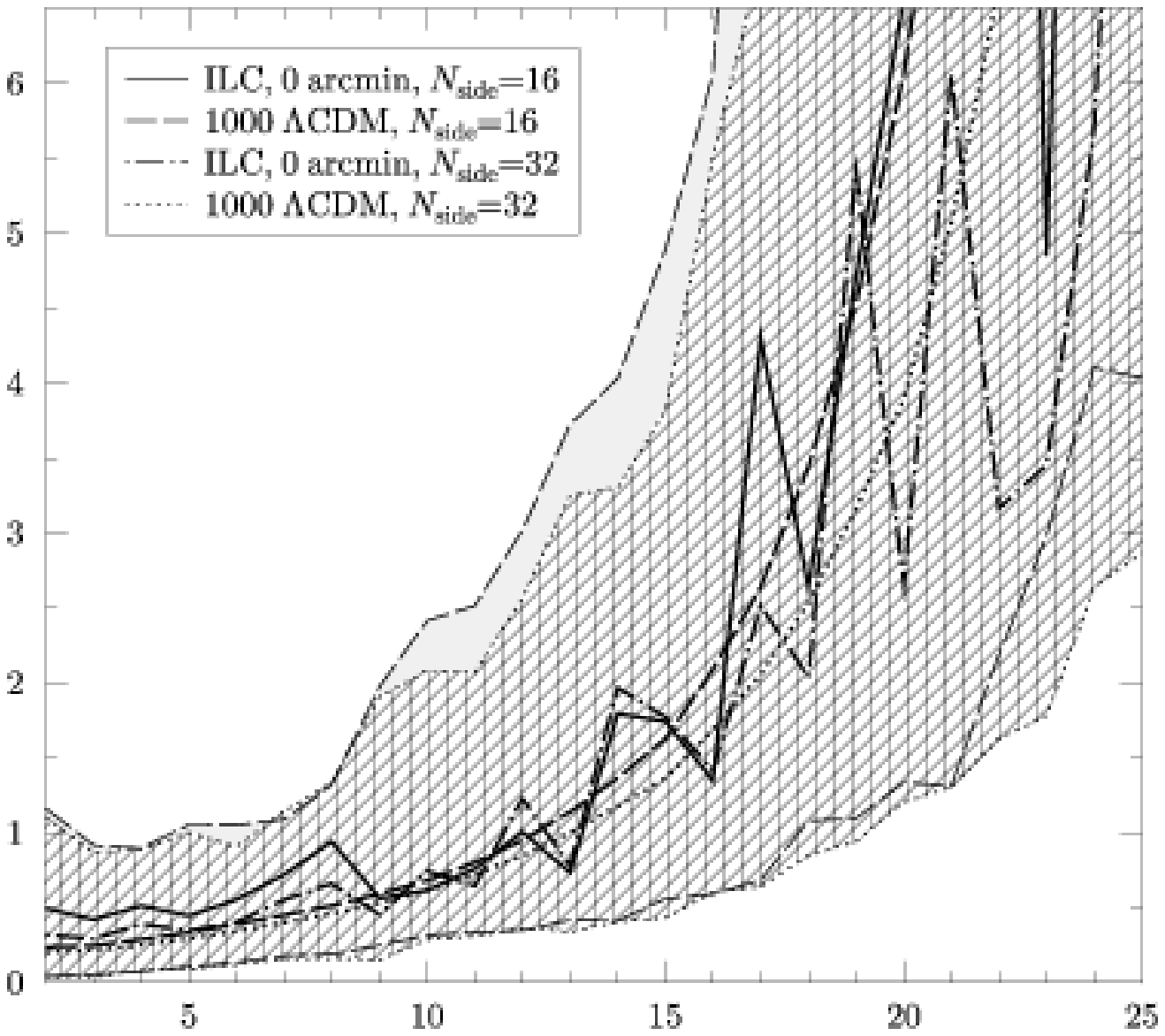}
\end{minipage}
\put(-325,65){$Q(l_{\hbox{\scriptsize max}})$}
\put(-84,-95){$l_{\hbox{\scriptsize max}}$}
\put(-275,38){(a) inside KQ85 mask}
}
\vspace*{-10pt}
{
\begin{minipage}{11cm}
\includegraphics[width=8.5cm]{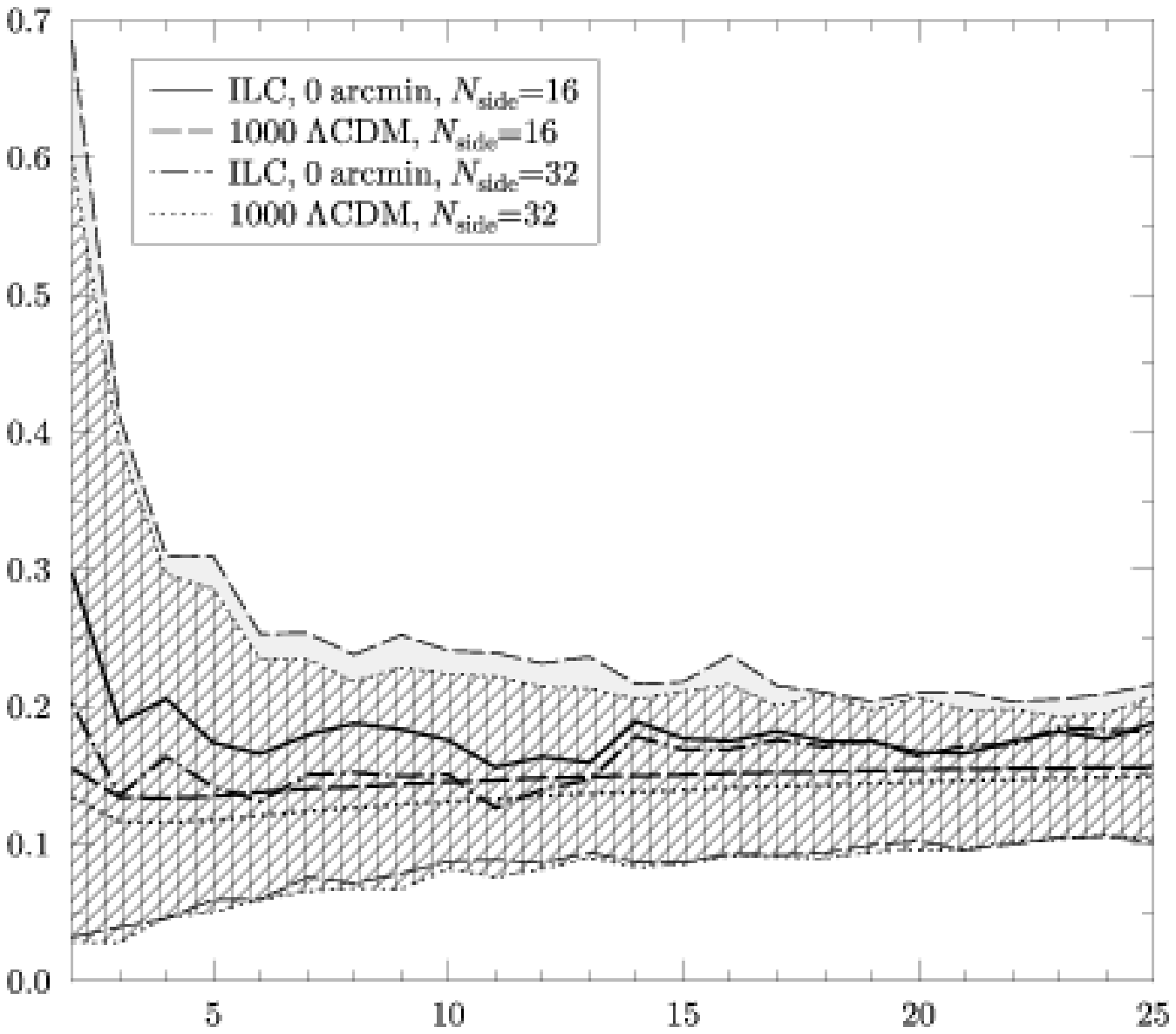}
\end{minipage}
\put(-325,75){$Q(l_{\hbox{\scriptsize max}})$}
\put(-84,-95){$l_{\hbox{\scriptsize max}}$}
\put(-275,38){(b) outside KQ85 mask}
}
\end{center}
\vspace*{-10pt}
\caption{\label{Fig:Q_ilc_und_1000lcdm_KQ85_FWHM_0arcmin_nside_16_und_32}
The normalised total error $Q(l_{\hbox{\scriptsize max}})$
is computed from the same data as in 
figure \ref{Fig:Q_ilc_und_1000lcdm_KQ85_FWHM_600arcmin_nside_16_und_32}.
The sole difference is that no additional smoothing is applied here,
which leads to larger values of $Q(l_{\hbox{\scriptsize max}})$ compared to
figure \ref{Fig:Q_ilc_und_1000lcdm_KQ85_FWHM_600arcmin_nside_16_und_32}.
Furthermore, a higher resolution leads only to a marginal improvement.
}
\end{figure}

The reconstruction error depends also on the resolution of the maps.
In order to address this point,
figure \ref{Fig:Q_ilc_und_1000lcdm_KQ85_FWHM_600arcmin_nside_16_und_32}
compares the total reconstruction errors $Q(l_{\hbox{\scriptsize max}})$
obtained from maps with HEALPix resolutions
$N_{\hbox{\scriptsize side}} = 16$ and 32.
The distribution of $Q(l_{\hbox{\scriptsize max}})$ of 1000 CMB simulations
as well as the curves belonging to the ILC maps are shown.
The maps are smoothed to 600 arcmin and the KQ85 (7yr) mask with
a mask threshold $x_{\hbox{\scriptsize th}}=0.5$ is used.
One observes that the reconstruction performs at the higher resolution 
$N_{\hbox{\scriptsize side}} = 32$ a bit better
than at the lower resolution $N_{\hbox{\scriptsize side}} = 16$.
This behaviour takes place within and outside the mask in a similar way.
As previously discussed, it is important to carry out the reconstruction
without a smoothing at intermediate steps.
For that reason,
figure \ref{Fig:Q_ilc_und_1000lcdm_KQ85_FWHM_0arcmin_nside_16_und_32}
shows the total reconstruction errors $Q(l_{\hbox{\scriptsize max}})$
without additional smoothing.
Aside from this difference the analysis is the same as in
figure \ref{Fig:Q_ilc_und_1000lcdm_KQ85_FWHM_600arcmin_nside_16_und_32}.
Omitting a smoothing procedure leads to an overall increase in
the total error $Q(l_{\hbox{\scriptsize max}})$.
In addition,
without smoothing the reconstruction at the higher resolution
$N_{\hbox{\scriptsize side}} = 32$ performs only marginally better
than at the lower resolution $N_{\hbox{\scriptsize side}} = 16$.
Thus, reconstructions of CMB maps at higher resolutions
$N_{\hbox{\scriptsize side}}$ do not yield a crucial improvement
if a smoothing has to be avoided.

\begin{figure}
\begin{center}
\vspace*{-15pt}
{
\begin{minipage}{11cm}
\includegraphics[width=8.5cm]{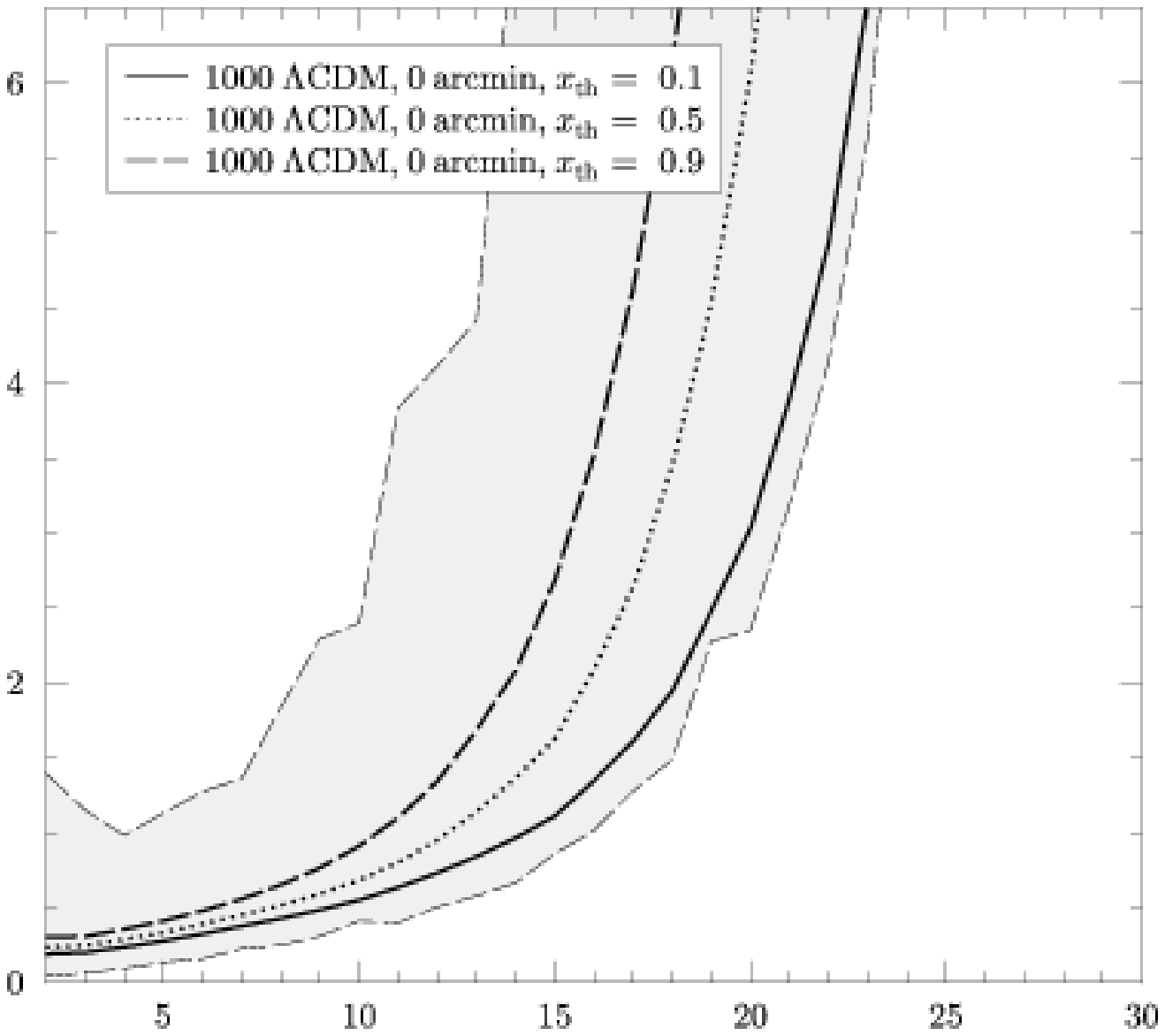}
\end{minipage}
\put(-325,65){$Q(l_{\hbox{\scriptsize max}})$}
\put(-84,-95){$l_{\hbox{\scriptsize max}}$}
\put(-275,40){(a) inside KQ85 mask}
}
\vspace*{-10pt}
{
\begin{minipage}{11cm}
\includegraphics[width=8.5cm]{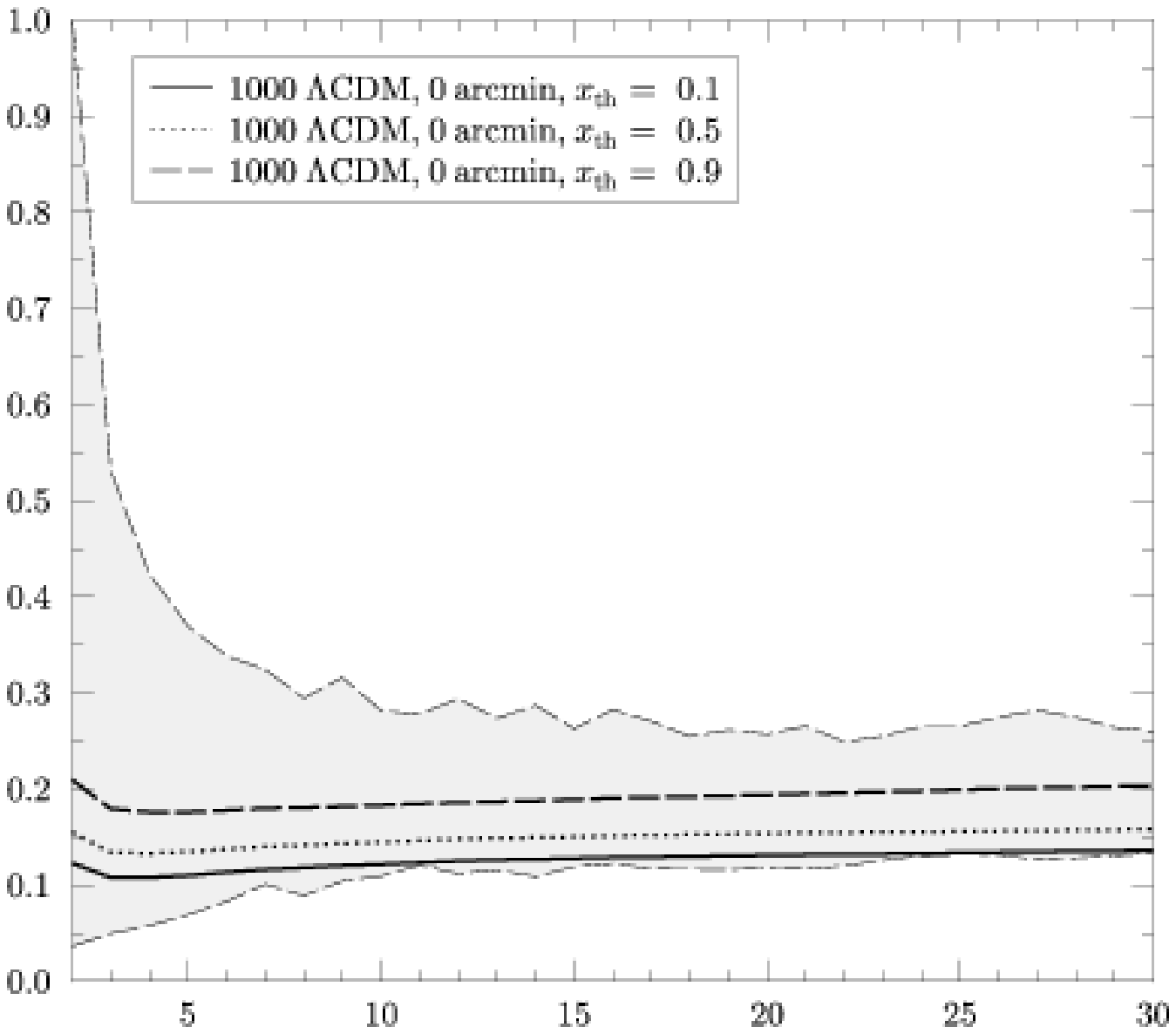}
\end{minipage}
\put(-325,78){$Q(l_{\hbox{\scriptsize max}})$}
\put(-84,-95){$l_{\hbox{\scriptsize max}}$}
\put(-275,40){(b) outside KQ85 mask}
}
\end{center}
\vspace*{-10pt}
\caption{\label{Fig:Q_1000lcdm_KQ85_FWHM_000arcmin_nside_16_abh_schwelle}
The mean value of the total reconstruction errors
$Q(l_{\hbox{\scriptsize max}})$
calculated from 1000 $\Lambda$CDM simulations
using the KQ85 (7yr) mask for the mask thresholds
$x_{\hbox{\scriptsize th}}=0.1$, 0.5 and 0.9 
at a resolution $N_{\hbox{\scriptsize side}} = 16$ are compared.
In the case of $x_{\hbox{\scriptsize th}}=0.9$ the distribution
of all reconstruction errors is given as a grey band.
The upper and the lower panel show the total reconstruction errors
computed within and outside the masks, respectively.
}
\end{figure}

Now we will discuss the influence of the mask threshold
$x_{\hbox{\scriptsize th}}$
defined in equation (\ref{Eq:not_masked_pixel_downgr}).
The lower the value of $x_{\hbox{\scriptsize th}}$,
the more pixels are excluded
in the neighbourhood of the boundary of a masked domain.
The total reconstruction errors $Q(l_{\hbox{\scriptsize max}})$ of 1000 CMB
simulations of the $\Lambda$CDM concordance model
for $x_{\hbox{\scriptsize th}}=0.1$, 0.5 and 0.9 
at a resolution $N_{\hbox{\scriptsize side}} = 16$
(without additional smoothing) are compared
in figure \ref{Fig:Q_1000lcdm_KQ85_FWHM_000arcmin_nside_16_abh_schwelle}.
For $x_{\hbox{\scriptsize th}}=0.1$, 0.5 and 0.9
the reconstruction method uses 
84.9\%, 80.8\% and 70.4\%  of the pixels of a map, respectively.
Thus the KQ85 (7yr) mask for $x_{\hbox{\scriptsize th}}=0.9$ at 
a resolution $N_{\hbox{\scriptsize side}} =16$ excludes a region 
almost as large as the initial KQ75 (7yr) mask,
and for $x_{\hbox{\scriptsize th}}=0.5$ the excluded region
is even smaller than the initial KQ85 (7yr) mask.
Since the reconstruction accuracy decreases by increasing the mask
and because a larger value of $x_{\hbox{\scriptsize th}}$ increases the mask,
the reconstruction has more difficulties for
larger $x_{\hbox{\scriptsize th}}$ as can be seen in figure 
\ref{Fig:Q_1000lcdm_KQ85_FWHM_000arcmin_nside_16_abh_schwelle}.
One observes in the upper panel
that the differences in the total reconstruction errors within the mask 
for the three thresholds $x_{\hbox{\scriptsize th}}$ are
relatively small as long as $l_{\hbox{\scriptsize max}} \lesssim 5$. 
But with increasing multipoles $l_{\hbox{\scriptsize max}}$
the total reconstruction error increases 
strongly with $x_{\hbox{\scriptsize th}}$.
The total reconstruction error within the mask for
$x_{\hbox{\scriptsize th}}=0.1$, 0.5 and 0.9
is typically larger than 1 for $l_{\hbox{\scriptsize max}} \gtrsim 15$, 13
and 11, respectively.
These represents the largest multipoles $l_{\hbox{\scriptsize max}}$
for which a reconstruction can be carried out
in order to avoid constructing random patterns.
But for cosmological parameter extractions,
even values of $Q(l_{\hbox{\scriptsize max}})$ near one can be insufficient,
since this implies an total error of a typical temperature fluctuation.

\begin{figure}
\begin{center}
\vspace*{-15pt}
{
\begin{minipage}{11cm}
\includegraphics[width=8.5cm]{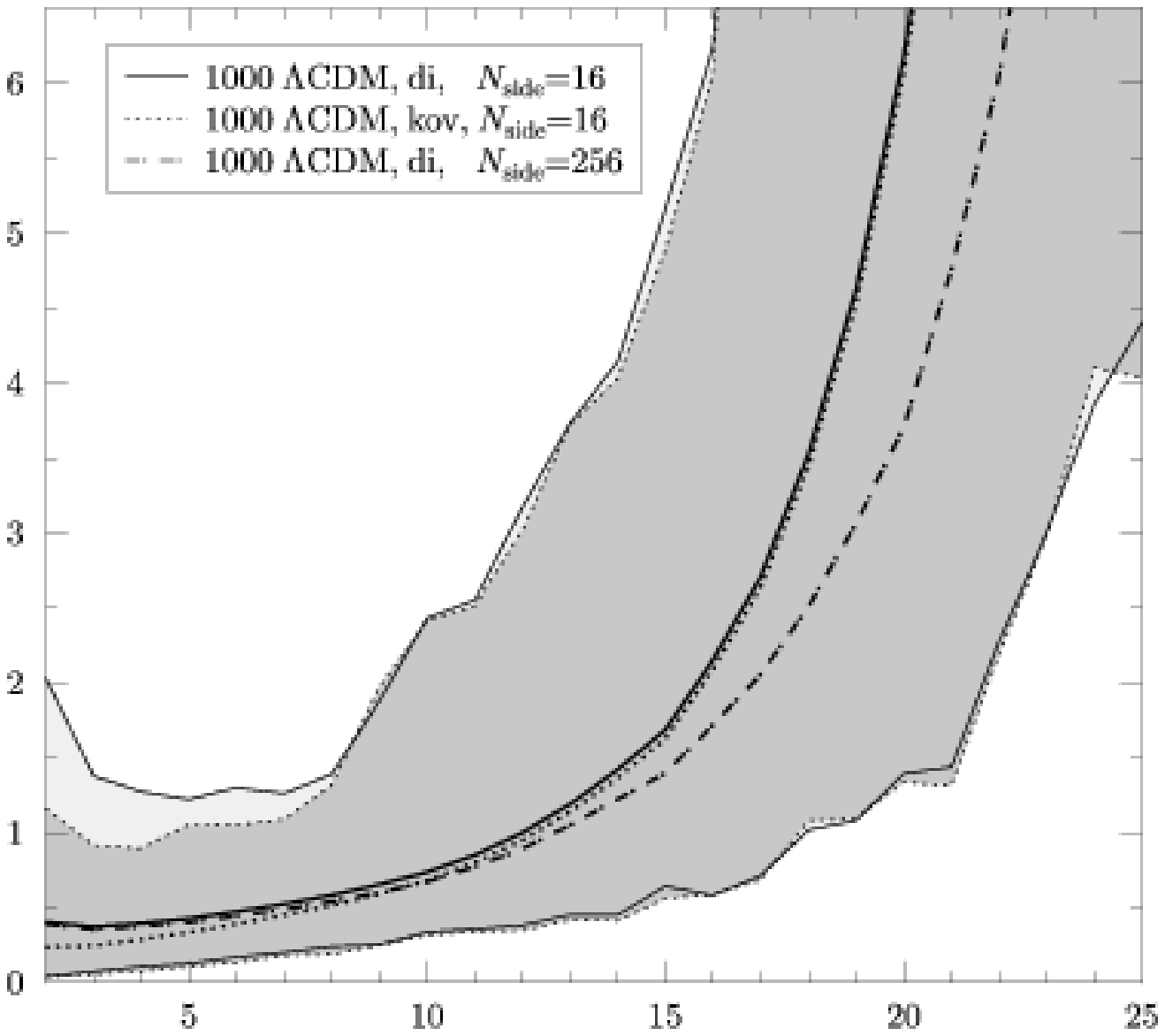}
\end{minipage}
\put(-325,65){$Q(l_{\hbox{\scriptsize max}})$}
\put(-84,-95){$l_{\hbox{\scriptsize max}}$}
\put(-275,40){(a) inside KQ85 mask}
}
\vspace*{-10pt}
{
\begin{minipage}{11cm}
\includegraphics[width=8.5cm]{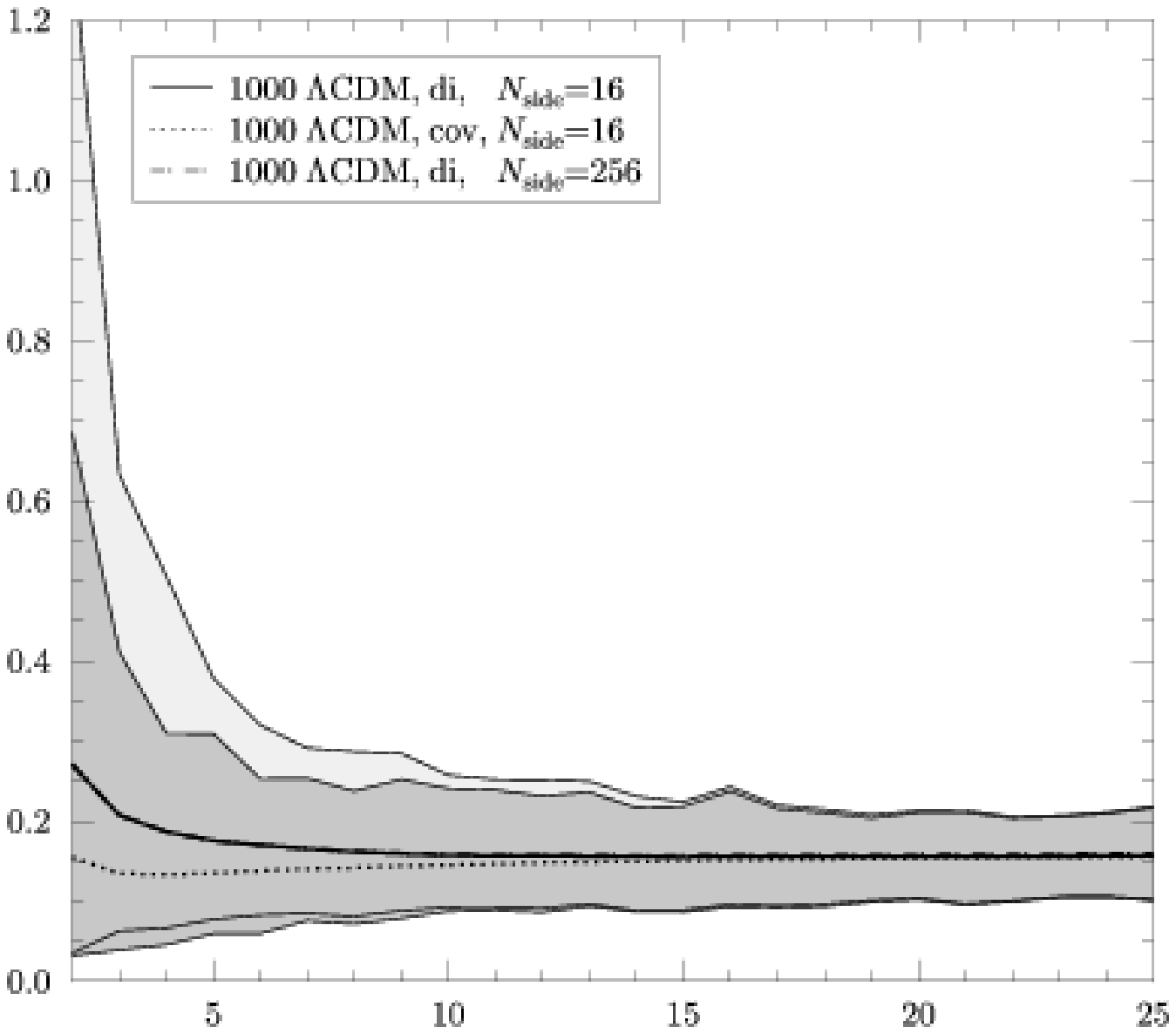}
\end{minipage}
\put(-325,75){$Q(l_{\hbox{\scriptsize max}})$}
\put(-84,-95){$l_{\hbox{\scriptsize max}}$}
\put(-275,40){(b) outside KQ85 mask}
}
\end{center}
\vspace*{-10pt}
\caption{\label{Fig:Q_1000lcdm_KQ85_FWHM_000arcmin_nside_16_bis_256_kov_und_di}
The total errors $Q(l_{\hbox{\scriptsize max}})$ of the reconstruction
using the covariance matrix (\ref{Eq:covariance_matrix}) (cov) are compared
with the total errors of the direct inversion (di).
Both methods use the KQ85 (7yr) mask with a mask threshold
$x_{\hbox{\scriptsize th}}=0.5$
at a resolution $N_{\hbox{\scriptsize side}} = 16$.
The direct inversion is also computed for $N_{\hbox{\scriptsize side}} = 256$.
The total error distribution of the 1000 CMB simulations is given as a
dark grey band for the algorithm using the covariance matrix
and as a light grey band for the direct inversion
($N_{\hbox{\scriptsize side}} = 16$).
In the upper panel the total reconstruction errors
$Q(l_{\hbox{\scriptsize max}})$ within the mask are given, and it is seen
that for $l_{\hbox{\scriptsize max}} \le 8$ the covariance method
possesses a smaller error distribution.
The lower panel shows the corresponding total reconstruction errors
outside the mask.
}
\end{figure}

Up to now, all reconstructions have used the
covariance matrix (\ref{Eq:covariance_matrix}) in equation (\ref{Eq:ar_by_A}).
Thus, we finally compare the reconstruction results
by using the covariance matrix with those of the method of direct inversion,
where $A$ is the unit matrix.
Choosing the unit matrix renders the method independent of an
underlying cosmological model, see equation (\ref{Eq:covariance_matrix}).
In figure \ref{Fig:Q_1000lcdm_KQ85_FWHM_000arcmin_nside_16_bis_256_kov_und_di} 
this comparison is carried out for the 1000 $\Lambda$CDM simulations using
the KQ85 (7yr) mask for a mask threshold $x_{\hbox{\scriptsize th}}=0.5$  
at a resolution of $N_{\hbox{\scriptsize side}} = 16$.
The reconstruction using the covariance matrix works better than
the method of direct inversion for $l_{\hbox{\scriptsize max}} \lesssim 8$.
For $l_{\hbox{\scriptsize max}} \gtrsim 9$ the difference between
the two methods is marginal.
Thus, for the reconstruction of the lowest multipoles
the algorithm using the covariance matrix is preferable,
whereas for the larger multipoles the direct inversion can be chosen
due to its smaller computational effort.
In the case of the method of direct inversion,
the mean value of the total reconstruction errors of 1000 simulations
at a resolution of $N_{\hbox{\scriptsize side}} = 256$ are also calculated,
which does not result in a crucial improvement
compared to the reconstruction at
a resolution of $N_{\hbox{\scriptsize side}} = 16$.
The similarity of both variants of the reconstruction algorithm
is due to the fact that the covariance matrix is already nearly diagonal.

The reconstructions of the 1000 $\Lambda$CDM simulations and  
of the ILC map are compared with
the multipole expansions of the full initial maps
up to the reconstruction value of $l_{\hbox{\scriptsize max}}$.
Thus, this comparison stringently needs the full sky map
in order to carry out the multipole expansion.
In the case of the simulations we know that the full initial maps
contain the so-called true CMB temperature fluctuations,
and the expansion can be done reliably.
The case of the ILC map is more involved
since the ILC map contains residual noise and foreground.
In particular, within and near the mask one would expect
remaining foreground to some extend.
The so-called true temperature fluctuations of the ILC map
are not necessarily the genuine CMB temperature fluctuations.
One has to keep this fact in mind,
when one tries to weight the quality of the ILC reconstruction
on the basis of the total reconstruction errors $Q(l_{\hbox{\scriptsize max}})$
of the simulations.
Thus it is possible that the reconstruction of the CMB of our Universe
from the ILC map is even worse than the result in figure 
\ref{Fig:Q_ilc_und_1000lcdm_KQ85_FWHM_0arcmin} suggests.

\section{Correlation function $C(\vartheta)$}

\begin{figure}
\begin{center}
\vspace*{-30pt}
{
\begin{minipage}{11cm}
\hspace*{-20pt}\includegraphics[width=10.0cm]{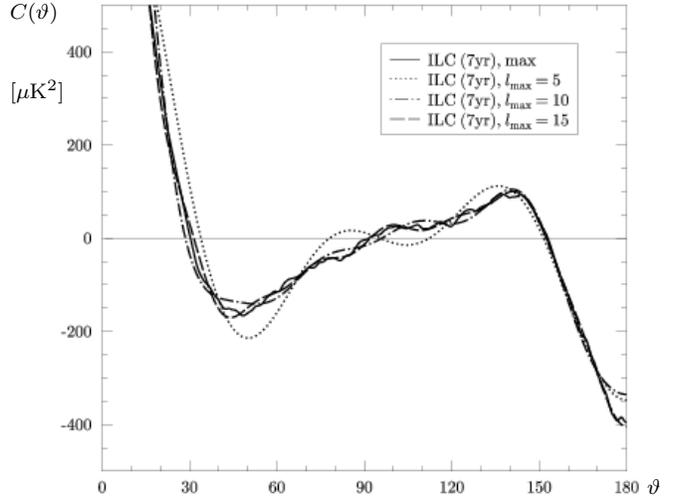}
\end{minipage}
\put(-325,85){$C(\vartheta)$}
\put(-325,55){[$\mu\hbox{K}^2$]}
\put(-84,-95){$\vartheta$}
}
\vspace*{-30pt}
\end{center}
\caption{\label{Fig:correlation_function_ilc_full_lmax_dependence}
The correlation functions of the full ILC map calculated from the multipoles 
up to $l_{\hbox{\scriptsize max}}=5$, 10 and 15 are compared with
the correlation functions of the full ILC map with maximal resolution (max).
}
\end{figure}

In this section we address the crucial question
whether the sky reconstructions can reliably be used
for the estimation of cosmological statistics.
The correlation function $C(\vartheta)$ is the suitable statistic
with respect to large angular scales $\vartheta$
which are relevant for the reconstruction algorithm.
This leads to the question of how sensitive
the correlation function $C(\vartheta)$,
equation (\ref{Eq:C_theta}), varies with respect to
the different sky reconstructions.
At first one has to determine the minimal value of the
reconstruction parameter $l_{\hbox{\scriptsize max}}$
which is needed for the reconstructed map in order to give
a correlation function $C(\vartheta)$ in agreement with the one
obtained from the initial map with full resolution.
Therefore, the 2-point correlation functions computed from the multipoles
up to $l_{\hbox{\scriptsize max}}=5$, 10 and 15 are compared with
the correlation function obtained from the full ILC map with maximal resolution
in figure \ref{Fig:correlation_function_ilc_full_lmax_dependence}.
From this figure one can read off that at least a reconstruction
parameter $l_{\hbox{\scriptsize max}} \gtrsim 10$ is needed
in order to have a good estimation of the correlation function at large scales.
The investigations of the previous section lead to the conclusion
that the KQ75 mask is already too large in order to allow
a reconstruction for such large values of $l_{\hbox{\scriptsize max}}$. 
For this reason the following computations are based all on
the KQ85 mask with a mask threshold $x_{\hbox{\scriptsize th}}=0.5$  
at a resolution of $N_{\hbox{\scriptsize side}} = 16$.
Furthermore, the reconstruction algorithm using the covariance matrix
is applied in the following,
since it gives for $l_{\hbox{\scriptsize max}} \gtrsim 10$
almost the same results as the direct inversion method.

\begin{figure}
\begin{center}
\vspace*{-12pt}
{
\begin{minipage}{11cm}
\includegraphics[width=8.2cm]{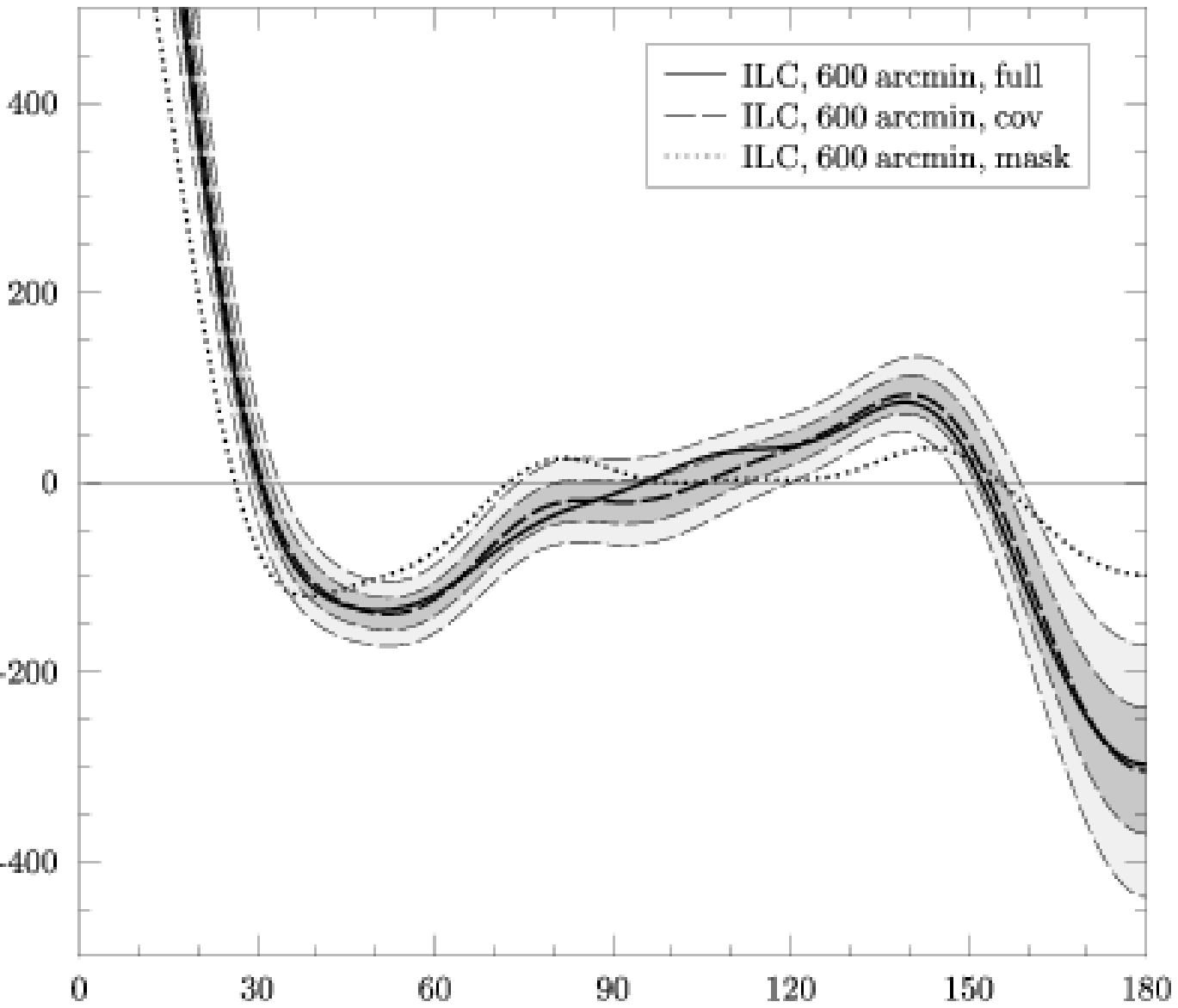}
\end{minipage}
\put(-325,85){$C(\vartheta)$}
\put(-325,55){[$\mu\hbox{K}^2$]}
\put(-84,-95){$\vartheta$}
\put(-275,-70){(a) $l_{\hbox{\scriptsize max}}=10$}
}
\vspace*{-5pt}
{
\begin{minipage}{11cm}
\includegraphics[width=8.2cm]{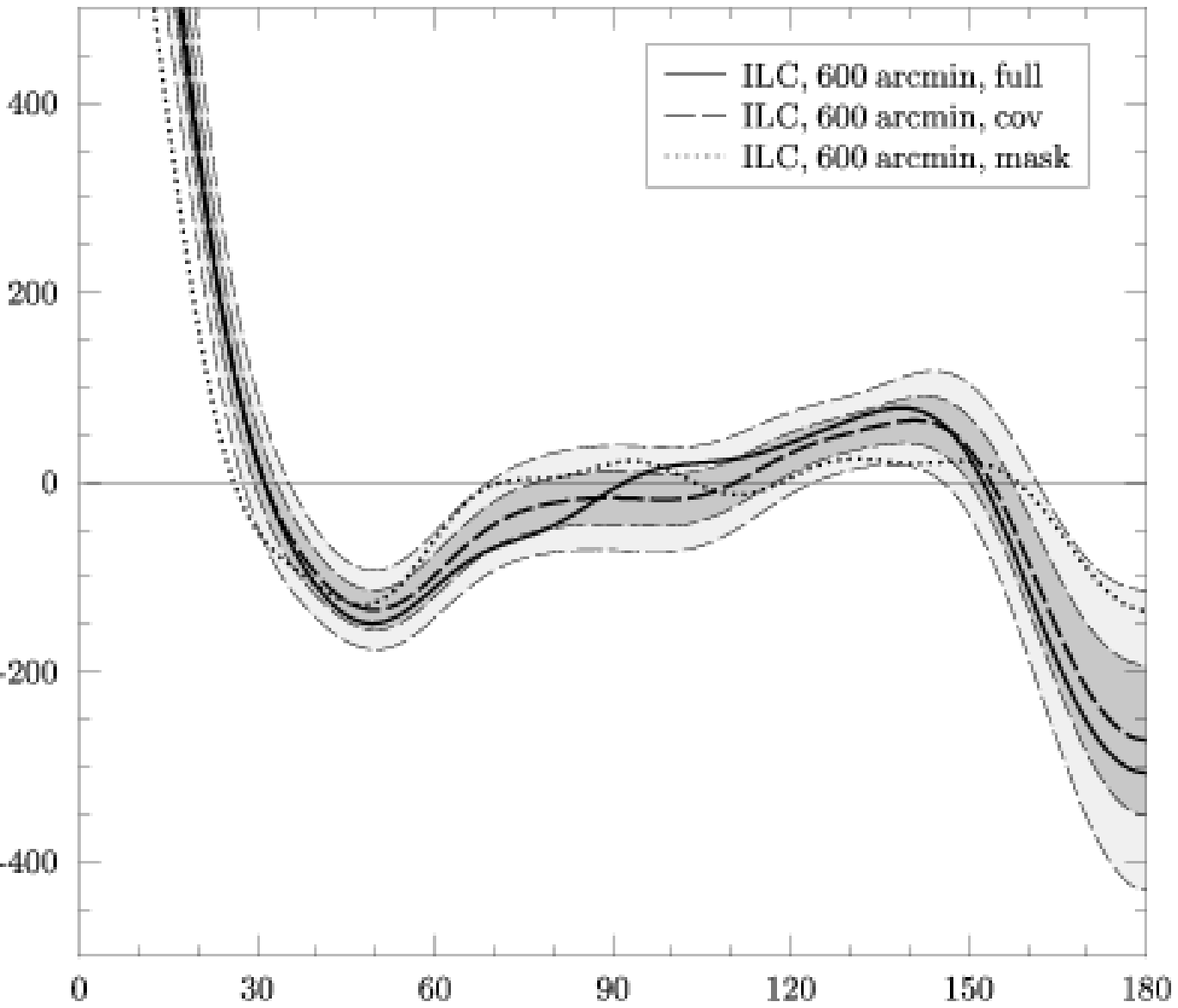}
\end{minipage}
\put(-325,85){$C(\vartheta)$}
\put(-325,55){[$\mu\hbox{K}^2$]}
\put(-84,-95){$\vartheta$}
\put(-275,-70){(b) $l_{\hbox{\scriptsize max}}=12$}
}
\vspace*{-5pt}
{
\begin{minipage}{11cm}
\includegraphics[width=8.2cm]{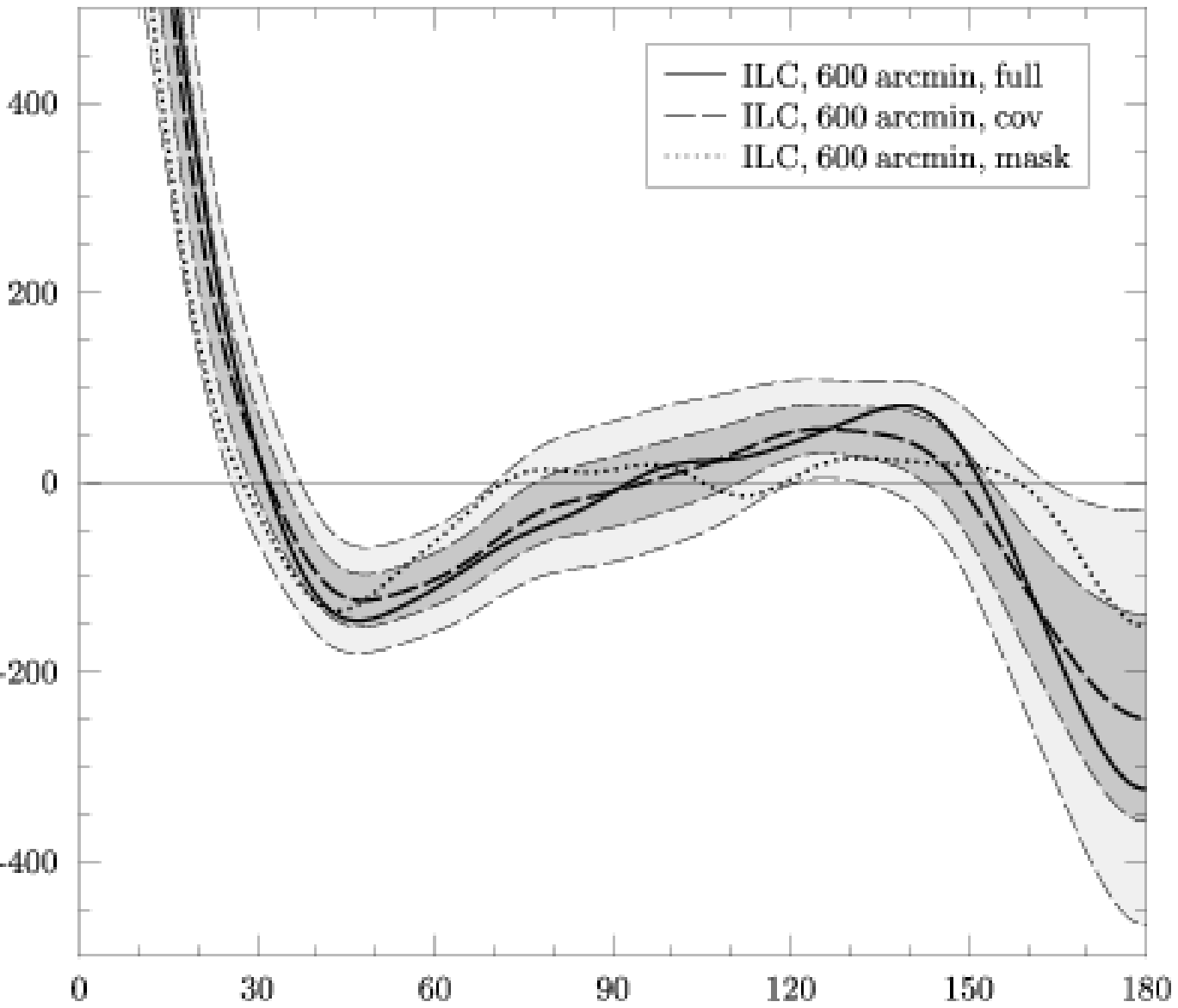}
\end{minipage}
\put(-325,85){$C(\vartheta)$}
\put(-325,55){[$\mu\hbox{K}^2$]}
\put(-84,-95){$\vartheta$}
\put(-275,-70){(c) $l_{\hbox{\scriptsize max}}=15$}
}
\vspace*{-10pt}
\end{center}
\caption{\label{Fig:correlation_function_ilc_KQ85_FWHM_600arcmin_nside_16}
In panel (a), (b) and (c) the correlation function $C(\vartheta)$
of   the reconstructed ILC map (cov) is compared 
with $C(\vartheta)$ of the full ILC map (full) and
$C(\vartheta)$ computed outside the mask (mask) 
for $l_{\hbox{\scriptsize max}}=10$, 12 and 15, respectively. 
Here the ILC map is additionally smoothed to 600 arcmin.
The reconstruction is based on the KQ85 mask
($x_{\hbox{\scriptsize th}}=0.5$ and $N_{\hbox{\scriptsize side}} = 16$).
The dark grey and light grey band show the 1$\sigma$ and 2$\sigma$ band
of $\sigma(\vartheta,l_{\hbox{\scriptsize max}})$,
eq.\,(\ref{Eq:error_C_theta_rec}),
calculated from 1000 $\Lambda$CDM simulations, respectively.
}
\end{figure}

In figure \ref{Fig:correlation_function_ilc_KQ85_FWHM_600arcmin_nside_16}
three correlation functions $C(\vartheta)$ are shown
which are calculated from the reconstructed ILC (7yr) map,
from the full ILC map and from the data outside the KQ85 mask.
The three panels show the results for the reconstruction parameters
$l_{\hbox{\scriptsize max}}=10$, 12 and 15.
Here the ILC (7yr) map with an additional smoothing of 600 arcmin is used.
The reconstructed correlation function is given with 1$\sigma$ and 2$\sigma$ 
deviations of the reconstruction errors which are estimated by 
\begin{equation}
\label{Eq:error_C_theta_rec}
\sigma(\vartheta,l_{\hbox{\scriptsize max}}) \; := \; 
\sqrt{
\frac{\sum_{k=1}^N(C_{\hbox{\scriptsize rec}}^{(k)}(\vartheta,l_{\hbox{\scriptsize max}})
-C_{\hbox{\scriptsize true}}^{(k)}(\vartheta,l_{\hbox{\scriptsize max}}))^2}
 {N}
}
\hspace{10pt} ,
\end{equation}
where the index $k$ counts the $N=1000$ CMB simulations
of the $\Lambda$CDM model.
This quantity takes into account only the error of the reconstruction of
pure CMB maps.
Additional contributions due to residual foreground and detector noise
in the ILC map propagate through the reconstruction algorithm
and are not considered here.
In all three cases ($l_{\hbox{\scriptsize max}}=10$,  12 and 15) the
correlation function of the reconstructed ILC map agrees well within the
$2\sigma$ errors with the correlation function of the full ILC map and less
with the correlation function computed outside the mask.
This observation is used by \cite{Efstathiou_Ma_Hanson_2009}
as an argument in favour of the statement
that the ``true'' correlation function is the one obtained
from the full ILC map.
However, as already emphasised,
the large additional smoothing of 600 arcmin transfers information
from the masked region into the data used by the algorithm.
For this reason this result is not surprising,
and the reconstructed correlation function is very questionable
since it is based on the pixels which should have been omitted.

\begin{figure}
\begin{center}
\vspace*{-12pt}
{
\begin{minipage}{11cm}
\includegraphics[width=8.2cm]{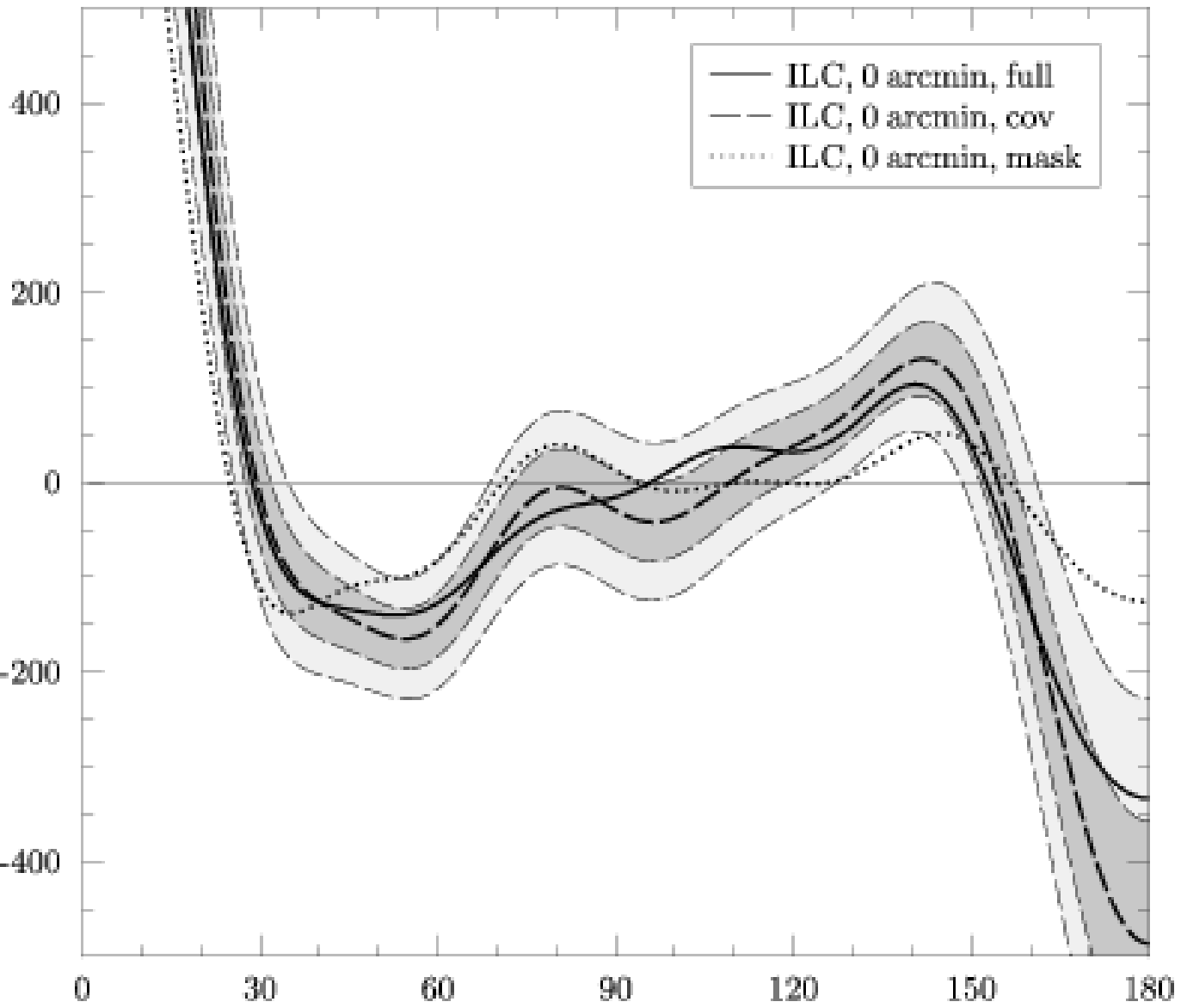}
\end{minipage}
\put(-325,85){$C(\vartheta)$}
\put(-325,55){[$\mu\hbox{K}^2$]}
\put(-84,-95){$\vartheta$}
\put(-275,-70){(a) $l_{\hbox{\scriptsize max}}=10$}
}
\vspace*{-5pt}
{
\begin{minipage}{11cm}
\includegraphics[width=8.2cm]{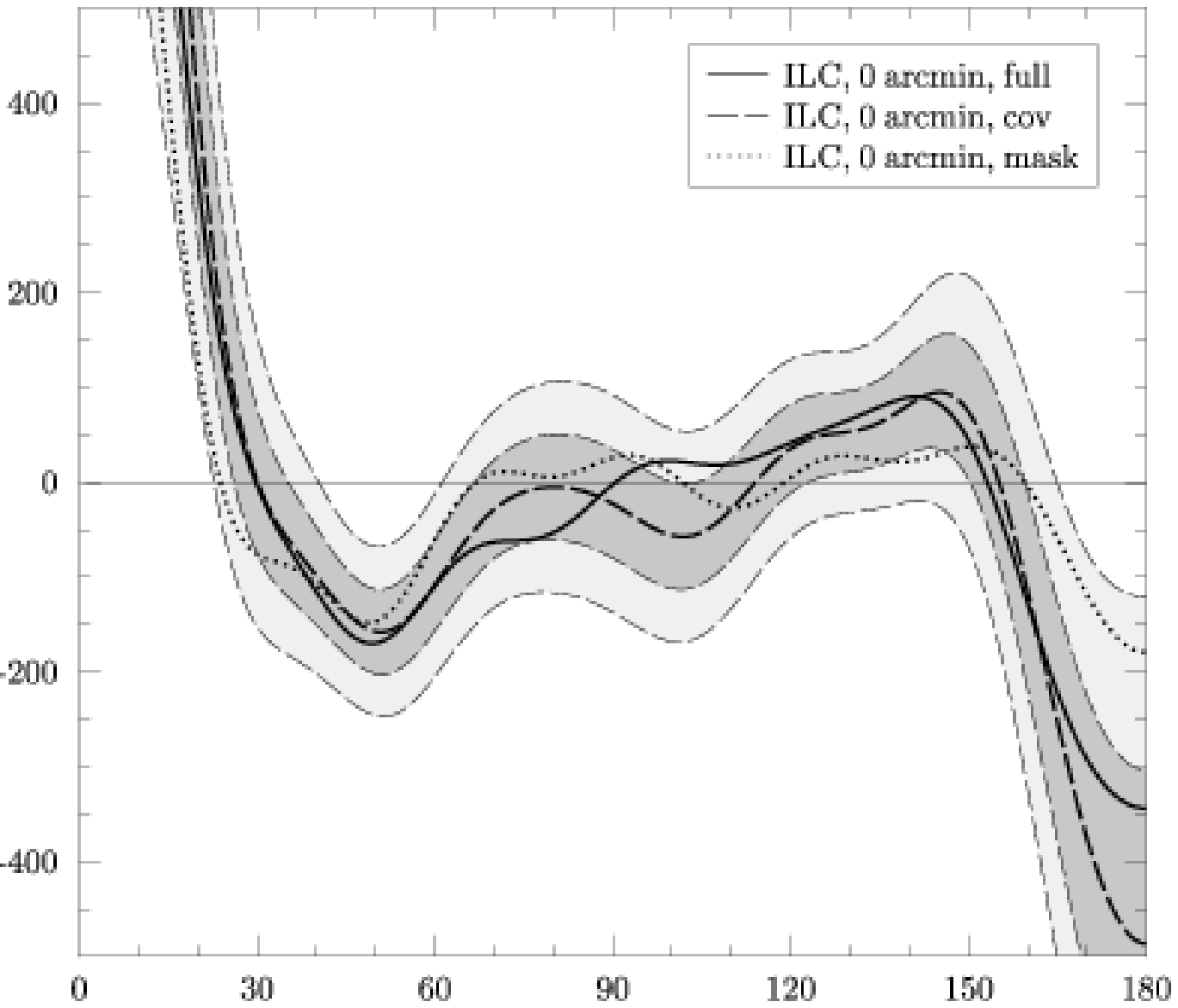}
\end{minipage}
\put(-325,85){$C(\vartheta)$}
\put(-325,55){[$\mu\hbox{K}^2$]}
\put(-84,-95){$\vartheta$}
\put(-275,-70){(b) $l_{\hbox{\scriptsize max}}=12$}
}
\vspace*{-5pt}
{
\begin{minipage}{11cm}
\includegraphics[width=8.2cm]{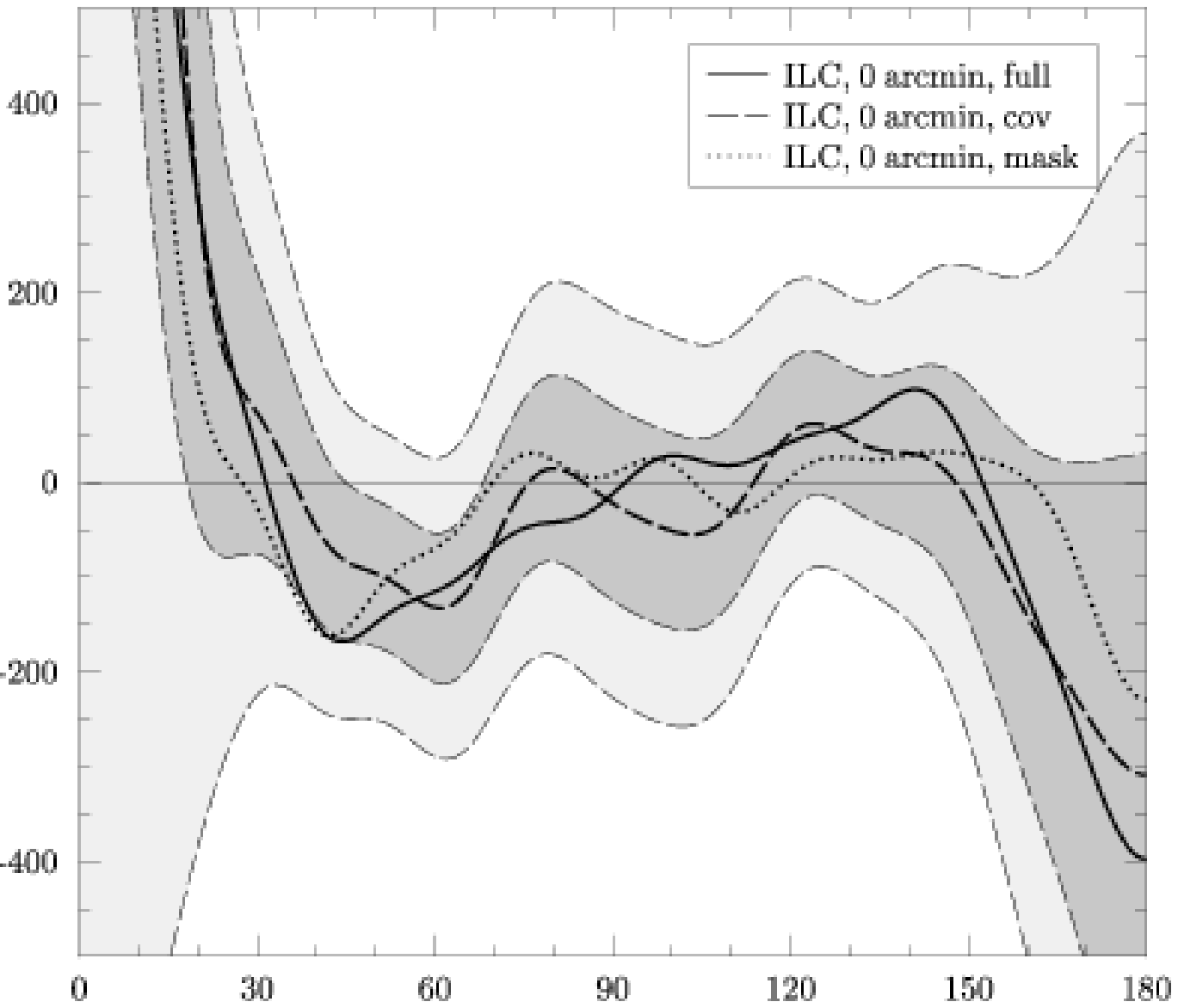}
\end{minipage}
\put(-325,85){$C(\vartheta)$}
\put(-325,55){[$\mu\hbox{K}^2$]}
\put(-84,-95){$\vartheta$}
\put(-275,-70){(c) $l_{\hbox{\scriptsize max}}=15$}
}
\vspace*{-10pt}
\end{center}
\caption{\label{Fig:correlation_function_ilc_KQ85_FWHM_000arcmin_nside_16}
The same computations are displayed as in
figure \ref{Fig:correlation_function_ilc_KQ85_FWHM_600arcmin_nside_16},
but here the ILC (7yr) map is used without additional smoothing
which leads to significantly larger errors.
}
\end{figure}

To avoid the problem of information transfer by smoothing 
the correlation function of the ILC (7yr) map
is also investigated without additional smoothing.
In figure \ref{Fig:correlation_function_ilc_KQ85_FWHM_000arcmin_nside_16}
the corresponding correlation functions are plotted.
The error (\ref{Eq:error_C_theta_rec}) is significantly larger now.
Without additional smoothing the differences between
the correlation function  of the reconstructed and of the full ILC (7yr) map
are larger, but both agree within the $2\sigma$ errors. 
This could lead to the conclusion that the reconstruction works
without additional smoothing and without a corresponding information transfer,
but for $l_{\hbox{\scriptsize max}}=10-12$ the errors are large
compared to the case with additional smoothing as it is shown
in figure \ref{Fig:correlation_function_ilc_KQ85_FWHM_600arcmin_nside_16}.
For $l_{\hbox{\scriptsize max}}\gtrsim 15$ the errors of
the reconstructed correlation function are too large to allow
any conclusion.
Due to these large errors the correlation function  
of the reconstructed ILC map also agrees on the same level with 
the correlation function obtained solely from the data outside the mask.
Thus, one cannot differentiate between these two cases.
The reconstructed correlation function shows sometimes 
a better agreement with the correlation function resulting from 
the masked ILC map and sometimes with the correlation function 
of the full ILC map.
The quality of this agreement depends 
on $\vartheta$ and $l_{\hbox{\scriptsize max}}$.
Thus, the reconstructed correlation function favours
neither the one of the masked nor the one of the full ILC map.
Due to the large errors the reconstructed correlation function
is uncertain by at least 100$\mu\hbox{K}^2$
and thus unsuited for the comparison with cosmological models.

\begin{figure}
\begin{center}
\vspace*{-40pt}
{
\begin{minipage}{11cm}
\hspace*{-20pt}\includegraphics[width=10.0cm]{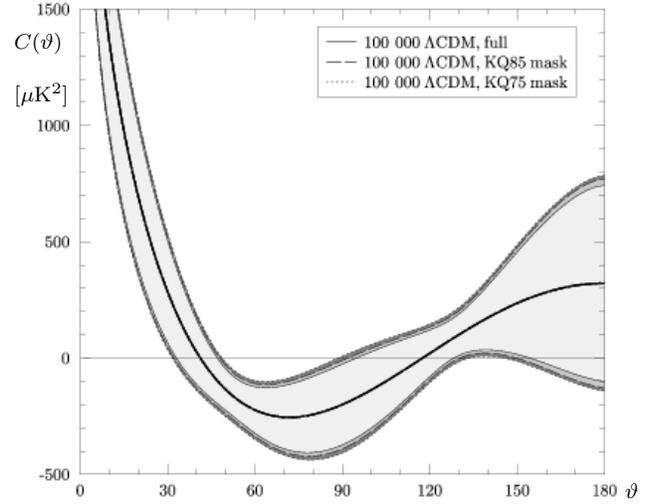}
\end{minipage}
\put(-315,75){$C(\vartheta)$}
\put(-315,55){[$\mu\hbox{K}^2$]}
\put(-84,-95){$\vartheta$}
}
\end{center}
\vspace*{-20pt}
\caption{\label{Fig:correlation_function_100000lcdm_modelle_full_KQ85_Kq75_nside_128_s0.5}
The mean value of the correlation function $C(\vartheta)$ of full maps
computed from 100\,000 CMB simulations of the $\Lambda$CDM model
is compared with the mean values of $C(\vartheta)$
using the KQ85 (7yr) mask and the KQ75 (7yr) mask.
The mean values cannot be distinguished in this plot.
All calculations are carried out with a resolution of
$N_{\hbox{\scriptsize side}} = 128$ and a mask threshold of
$x_{\hbox{\scriptsize th}}=0.5$.
The light grey band gives the 1$\sigma$ standard deviation
calculated from full maps of the 100\,000 CMB simulations.
The medium grey and the dark grey band correspond to
the 1$\sigma$ standard deviations calculated from the same 100\,000 maps,
but now outside the KQ85 and outside the KQ75 mask, respectively.
}
\end{figure}

The above discussion puts forward arguments 
against the reconstruction method and favours methods
which use only the data outside a given mask.
To provide a firm footing for the latter,
the influence of the KQ85 and KQ75 masks onto the ensemble average and
the cosmic variance of the correlation function $C(\vartheta)$
is now investigated with respect to the $\Lambda$CDM model.
Both quantities are shown in figure 
\ref{Fig:correlation_function_100000lcdm_modelle_full_KQ85_Kq75_nside_128_s0.5}
for an ensemble of 100\,000 CMB simulations of the $\Lambda$CDM model.
The ensemble average and the standard deviation is computed from
the correlation functions $C(\vartheta)$ obtained from
the data of full maps, outside the KQ85 mask and outside the KQ75 mask.
A resolution of $N_{\hbox{\scriptsize side}} = 128$ and
a mask threshold $x_{\hbox{\scriptsize th}}=0.5$ is used. 
The ensemble averages are identical in all three cases.
The standard deviation slightly increases with the size of the mask.
The smallest 1$\sigma$ standard deviation is obtained by using no mask at all,
the next larger deviation belongs to the KQ85 mask
whereas the largest deviation is due to the larger KQ75 mask.
The increase of the standard deviation for the masked data is, however,
small compared to the uncertainty resulting from the reconstruction method
for the same mask.
This is a further argument for having more confidence in the
correlation function $C(\vartheta)$ computed from the data outside the mask
than in the one obtained from a reconstructed full map.

\section{The large angular scale power}

\begin{figure}
\begin{center}
\vspace*{-15pt}
{
\begin{minipage}{11cm}
\hspace*{-10pt}\includegraphics[width=9.0cm]{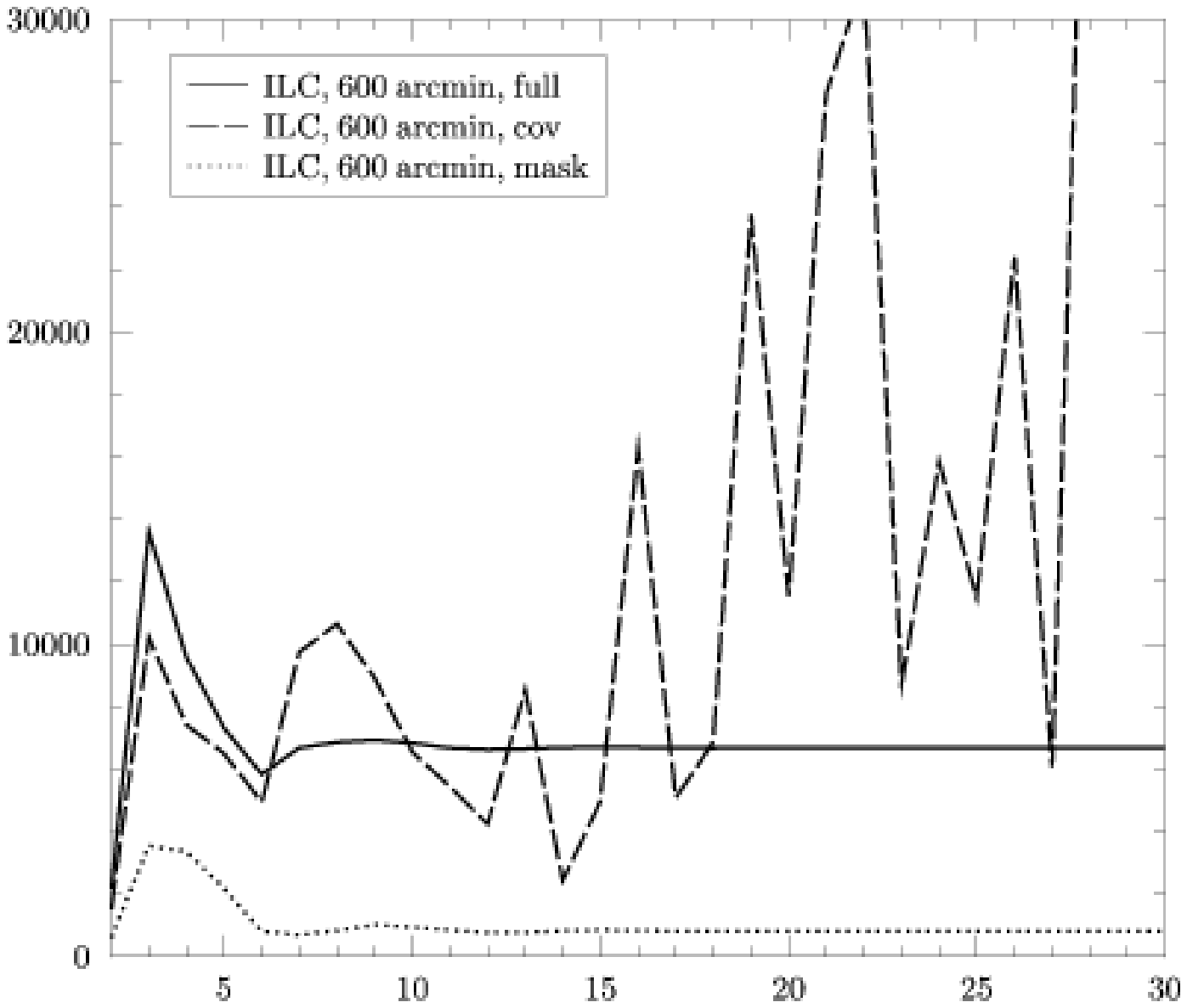}
\end{minipage}
\put(-325,65){$S(60^{\circ})$}
\put(-325,45){[$\mu\hbox{K}^4$]}
\put(-84,-95){$l_{\hbox{\scriptsize max}}$}
\put(-265,40){(a)}
}
\vspace*{-15pt}
{
\begin{minipage}{11cm}
\hspace*{-10pt}\includegraphics[width=9.0cm]{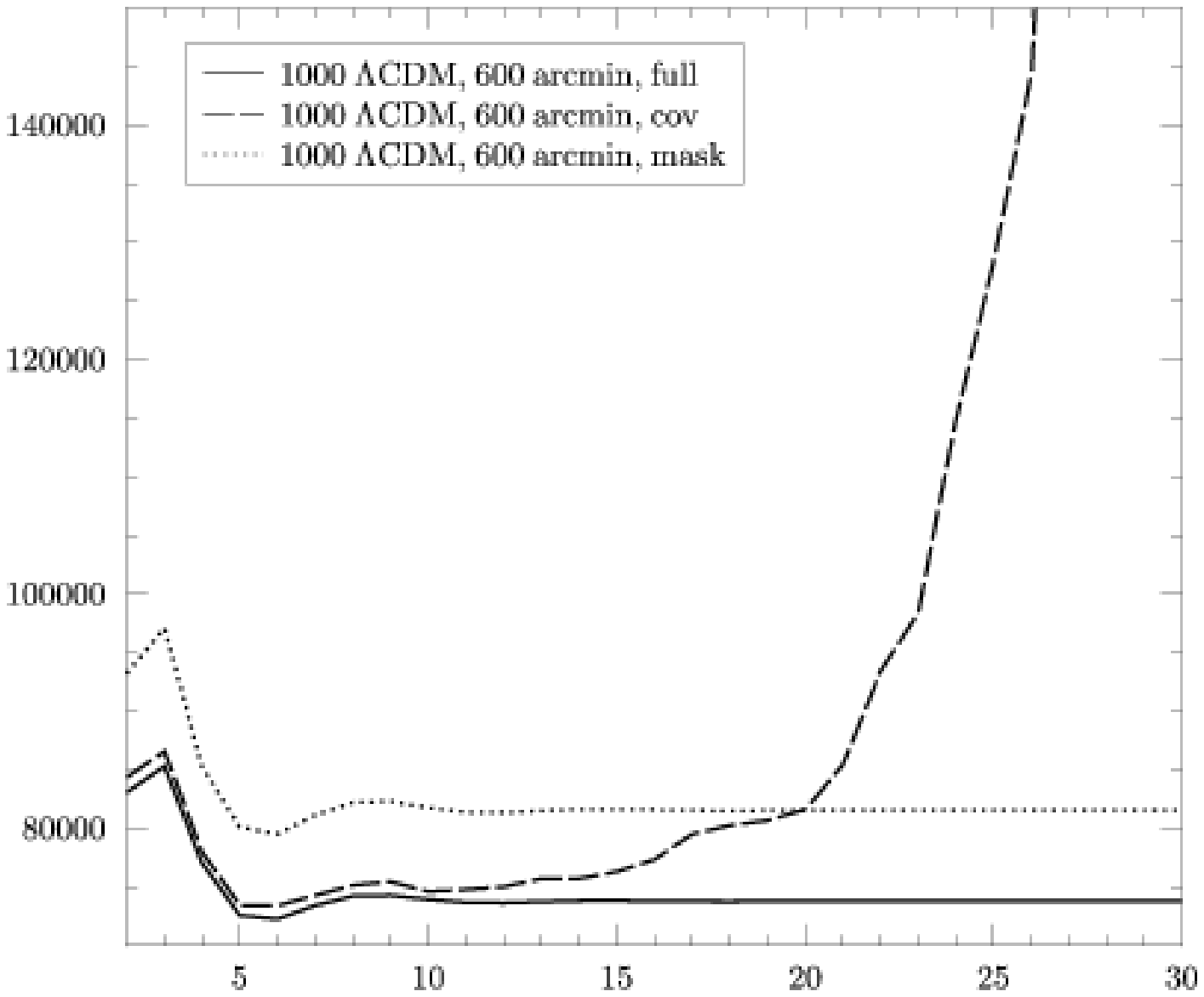}
\end{minipage}
\put(-325,85){$S(60^{\circ})$}
\put(-325,55){[$\mu\hbox{K}^4$]}
\put(-84,-95){$l_{\hbox{\scriptsize max}}$}
\put(-265,40){(b)}
}
\end{center}
\vspace*{-15pt}
\caption{\label{Fig:s_statistik_1000lcdm_und_ilc_KQ85_FWHM_600arcmin_nside_16}
In panel (a) the $S(60^{\circ})$ statistic of  
the reconstructed ILC (7yr) map (cov), the full ILC (7yr) map (full) 
and the  ILC (7yr) map outside the applied mask (mask) is 
plotted as a function of $l_{\hbox{\scriptsize max}}$. 
Here the ILC map is additionally smoothed with 600 arcmin.
The reconstruction is carried out applying the KQ85 (7yr) mask
for the mask threshold $x_{\hbox{\scriptsize th}}=0.5$ 
using the resolution $N_{\hbox{\scriptsize side}} = 16$.
In panel (b) the mean values of the $S(60^{\circ})$
statistic calculated from 1000 CMB simulations of the $\Lambda$CDM model
are plotted. 
}
\end{figure}

The main difference between the ILC correlation functions $C(\vartheta)$
computed from the data outside the KQ85 mask and from all pixels is
that the former displays much less power at angular scales above
$60^{\circ}$ than the latter.
To describe this observation the $S(60^{\circ})$ statistic
\begin{equation}
\label{Eq:S_statistic_60}
S(60^{\circ})\; := \; \int^{\cos(60^{\circ})}_{-1}
d\cos\vartheta \; |C(\vartheta)|^2 
\hspace{10pt} 
\end{equation}
has been introduced \citep{Spergel_et_al_2003},
which measures the power on scales larger than $60^{\circ}$.
In the following the $S(60^{\circ})$ statistic is discussed as a function
of the reconstruction parameter $l_{\hbox{\scriptsize max}}$.
At first the results obtained from maps with an additional smoothing
of 600 arcmin are presented.
In figure \ref{Fig:s_statistik_1000lcdm_und_ilc_KQ85_FWHM_600arcmin_nside_16}a
the $S(60^{\circ})$ statistic is computed from the full ILC map,
from the data outside the mask and from the reconstructed ILC map.
It is striking to see that in the case of the reconstructed map
the $S(60^{\circ})$ statistic does not converge to a stable value
as it is the case for the full ILC map as well as for the masked ILC map
for $l_{\hbox{\scriptsize max}} \gtrsim 6$.
This emphasises the uncertainty of reconstructed maps despite 
the information transfer by smoothing the map. 
In figure \ref{Fig:s_statistik_1000lcdm_und_ilc_KQ85_FWHM_600arcmin_nside_16}b
the mean value of the $S(60^{\circ})$ statistic,
calculated from 1000 CMB simulations of the $\Lambda$CDM model,
is displayed again with an additional smoothing of 600 arcmin.
As in figure
\ref{Fig:s_statistik_1000lcdm_und_ilc_KQ85_FWHM_600arcmin_nside_16}a
the statistic is plotted for full maps, for reconstructed maps and
for masked maps.
Since the $\Lambda$CDM model has significantly more power at scales
above $\vartheta=60^{\circ}$, the scale of the ordinate now differs
from that of figure 
\ref{Fig:s_statistik_1000lcdm_und_ilc_KQ85_FWHM_600arcmin_nside_16}a.
This should be noted by comparing both panels.
For $l_{\hbox{\scriptsize max}} \lesssim 10$ the mean value of 
the $S(60^{\circ})$ statistic of the reconstructed maps differs only
marginally from that of the full maps, but for
$l_{\hbox{\scriptsize max}} \gtrsim 10$ the differences are pronounced.
If the reconstruction method would work perfectly,
the mean values of the $S(60^{\circ})$ statistic
of reconstructed maps and of full maps should be identical, 
but there is a systematic overestimation of power in the reconstructed maps.
This clearly hints to a systematic error by using the reconstruction method. 
The overestimation of the mean value of the $S(60^{\circ})$ statistic 
obtained from 1000 simulations again stresses the problems 
of the reconstruction method for $l_{\hbox{\scriptsize max}} \gtrsim 10$.

\begin{figure}
\begin{center}
\vspace*{-15pt}
{
\begin{minipage}{11cm}
\hspace*{-10pt}\includegraphics[width=9.0cm]{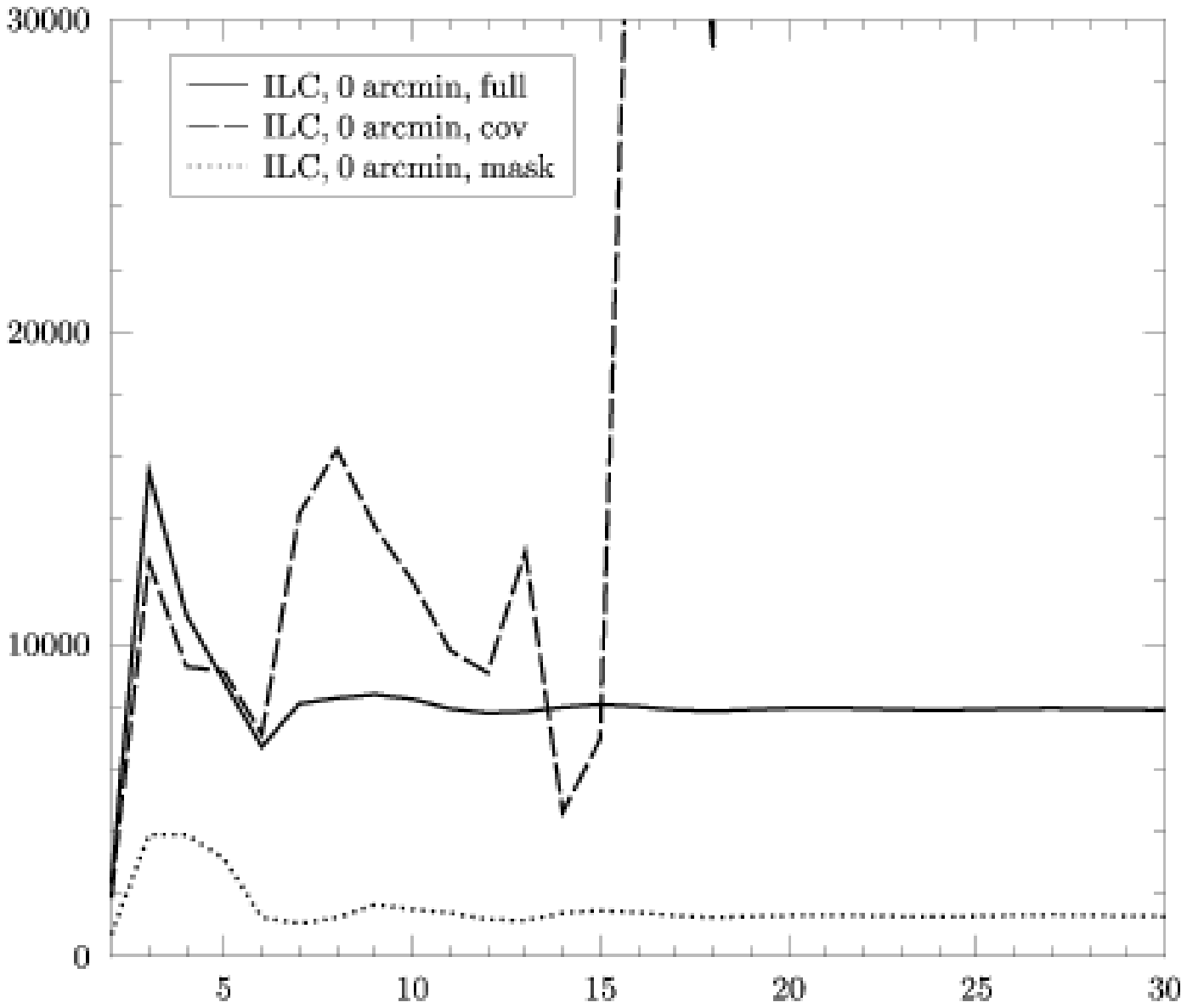}
\end{minipage}
\put(-325,65){$S(60^{\circ})$}
\put(-325,45){[$\mu\hbox{K}^4$]}
\put(-84,-95){$l_{\hbox{\scriptsize max}}$}
\put(-265,40){(a)}
}
\vspace*{-15pt}
{
\begin{minipage}{11cm}
\hspace*{-10pt}\includegraphics[width=9.0cm]{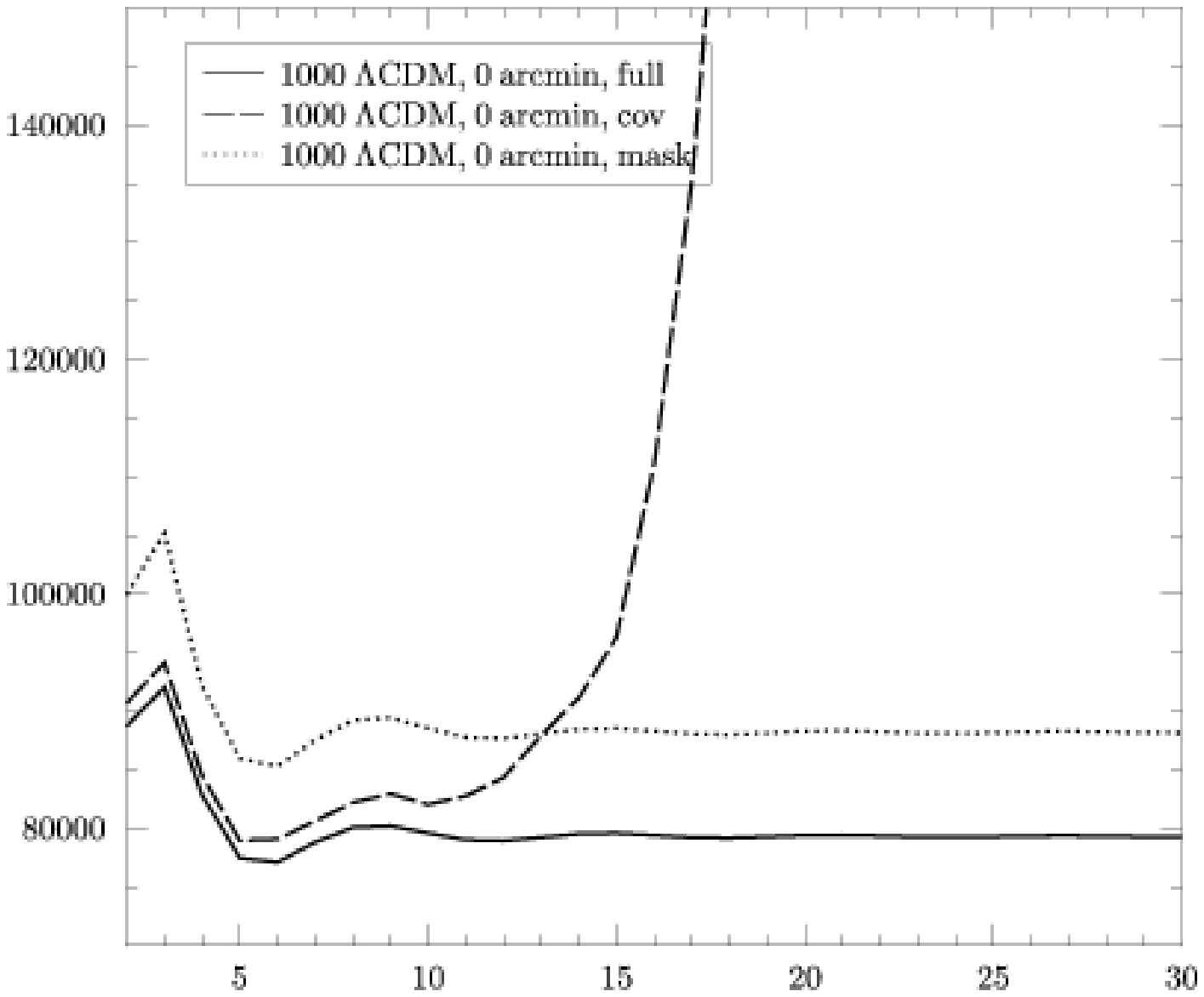}
\end{minipage}
\put(-325,85){$S(60^{\circ})$}
\put(-325,55){[$\mu\hbox{K}^4$]}
\put(-84,-95){$l_{\hbox{\scriptsize max}}$}
\put(-265,40){(b)}
}
\end{center}
\vspace*{-15pt}
\caption{\label{Fig:s_statistik_1000lcdm_und_ilc_KQ85_FWHM_000arcmin_nside_16}
The same quantities as in
figure \ref{Fig:s_statistik_1000lcdm_und_ilc_KQ85_FWHM_600arcmin_nside_16}
are displayed for the ILC (7yr) map in panel (a)
and for 1000 $\Lambda$CDM simulations in panel (b), 
but now they are calculated from the maps without additional smoothing.
}
\end{figure}

Furthermore, figure 
\ref{Fig:s_statistik_1000lcdm_und_ilc_KQ85_FWHM_600arcmin_nside_16}b
reveals that the mean value of the $S(60^{\circ})$ statistic increases by
10\% by using only the data outside the KQ85 mask compared
to the full pixel input.
It is interesting to note
that the reverse behaviour occurs in the case of the ILC map.
There, the $S(60^{\circ})$ statistic obtained from the masked map
is below that of the full map.
Using the above observation based on 1000 simulations,
one would rather expect that the $S(60^{\circ})$ statistic of the
full ILC map should lie 10\% below the one of the masked map.
The ILC map displays in this respect an untypical behaviour.

In figure \ref{Fig:s_statistik_1000lcdm_und_ilc_KQ85_FWHM_000arcmin_nside_16}
the same quantities are displayed as 
in figure \ref{Fig:s_statistik_1000lcdm_und_ilc_KQ85_FWHM_600arcmin_nside_16},
but now calculated without additional smoothing of the maps.
Compared to the previous case with smoothing,
the $S(60^{\circ})$ statistic of the reconstructed ILC map fluctuates
even stronger for $l_{\hbox{\scriptsize max}} \gtrsim 6$
which emphasises the strong uncertainty of the reconstruction without
the information transfer caused by smoothing. 
For $l_{\hbox{\scriptsize max}} > 15$ the $S(60^{\circ})$ statistic
of the reconstructed ILC (7yr) map yields extremely large values
which corresponds to the overestimation of the power
of the correlation function at large scales. 
The mean value of the $S(60^{\circ})$ statistic 
calculated from 1000 CMB simulations of the $\Lambda$CDM model
is plotted in
figure \ref{Fig:s_statistik_1000lcdm_und_ilc_KQ85_FWHM_000arcmin_nside_16}b. 
Again the overestimation of power in the case of reconstructed maps
is much more pronounced as in the case without an additional smoothing.
Now this overestimation increases very fast already
for $l_{\hbox{\scriptsize max}} \gtrsim 12$
which shows that for these values of $l_{\hbox{\scriptsize max}}$
the reconstructed results definitely contain significant systematic errors.
For individual maps the errors can be significant for even lower values of
$l_{\hbox{\scriptsize max}}$ as shown in figure
\ref{Fig:s_statistik_1000lcdm_und_ilc_KQ85_FWHM_000arcmin_nside_16}a
for the ILC map.

The figures \ref{Fig:s_statistik_1000lcdm_und_ilc_KQ85_FWHM_600arcmin_nside_16}
and \ref{Fig:s_statistik_1000lcdm_und_ilc_KQ85_FWHM_000arcmin_nside_16}
also reveal the fact that the low value of $S(60^{\circ})$ of the ILC map
is rather unusual.
As can be read off from the figures this value is much smaller than the
mean value due to the 1000 $\Lambda$CDM simulations.
The $S(60^{\circ})$ statistic of the ILC map is now compared
with those of 100\,000 $\Lambda$CDM simulations
in order to find the frequency of simulations with even lower power
than contained in the ILC map, i.\,e.\ which satisfy the condition
\citep{Spergel_et_al_2003,Copi_Huterer_Schwarz_Starkman_2010}
$$
S_{\scriptsize \Lambda \hbox{CDM}}(60^{\circ}) \; < \;
S_{\hbox{\scriptsize ILC}}(60^{\circ})
\hspace{10pt} .
$$
This condition is satisfied by only 5244, 54 and 16 models out of the
100\,000 CMB simulations
where $S(60^{\circ})$ is computed from data of the full map,
outside the KQ85 mask, and outside the KQ75 mask, respectively.
As discussed above the values of $S(60^{\circ})$ of the ILC map
is lower for the masked map than for the full map which is unusual.
How unusual this behaviour is, becomes evident by the fact that
none of the 100\,000 $\Lambda$CDM simulations simultaneously satisfy
the conditions 
$S_{\scriptsize \Lambda \hbox{CDM}}(60^{\circ}) >
S_{\hbox{\scriptsize ILC}}(60^{\circ})$ 
for data of the full map and  
$S_{\scriptsize \Lambda \hbox{CDM}}(60^{\circ}) <
S_{\hbox{\scriptsize ILC}}(60^{\circ})$
for data outside the KQ85 mask.
Using instead of the KQ85 mask the KQ75 mask leads to only 2 models
out of 100\,000 CMB simulations which satisfy both conditions simultaneously.
This is a clear hint that the ILC map shows an untypical behaviour
compared to the $\Lambda$CDM concordance model.

\section{Summary}

In this paper we have investigated the reconstruction of pixel values
within masks using inversion methods.
We have compared the algorithm using the covariance matrix
with the direct inversion method.
The comparison shows that for the reconstruction of multipoles
$l_{\hbox{\scriptsize max}}\lesssim 8$
the algorithm using the covariance matrix gives slightly smaller errors
than the direct inversion method.
For this reason we focus on the reconstruction method 
which uses the covariance matrix. 

An important point of our analysis is that the additional smoothing of
$10^{\circ}$ of the ILC map in \cite{Efstathiou_Ma_Hanson_2009} 
transfers information from pixels inside the mask to pixels outside.
This should not happen since the pixel values inside the mask are
considered as insecure and should be ignored.
The smoothing, however, leads to a quantitatively better reconstruction
of the ILC map,
if the masked domain is considered as containing the true pixel values
which have to be recovered by the reconstruction.
But the contribution of information from masked pixels makes a 
reconstructed map obtained from data with additional smoothing unacceptable
in a realistic application.

The errors of the pixel values of the reconstructed map are compared
with the mean temperature fluctuation
$\sigma_{\hbox{\scriptsize true}}(l_{\hbox{\scriptsize max}})$
defined in equation (\ref{Eq:Ortsraum_Normierung}).
In the case of the KQ75 (7yr) mask the reconstructed pixels inside the mask
have errors larger than
$\sigma_{\hbox{\scriptsize true}}(l_{\hbox{\scriptsize max}})$
already for $l_{\hbox{\scriptsize max}} \gtrsim 6$
such that the reconstruction is unusable as revealed by
figure \ref{Fig:Q_ilc_KQ75}.
Even for smaller values of $l_{\hbox{\scriptsize max}}$
the errors are at least of the order of
$\frac 12\sigma_{\hbox{\scriptsize true}}(l_{\hbox{\scriptsize max}})$.
The culprit for this negative result is the size of the KQ75 (7yr) mask
being too large for a reconstruction.

In the case of the smaller KQ85 (7yr) mask the reconstruction
without additional smoothing leads
for $l_{\hbox{\scriptsize max}} \gtrsim 13$
to errors larger than 
$\sigma_{\hbox{\scriptsize true}}(l_{\hbox{\scriptsize max}})$
such that a rough reconstruction can be done for
$l_{\hbox{\scriptsize max}}$ around 10.
However, even with this restriction, errors larger than
$\frac 12\sigma_{\hbox{\scriptsize true}}(l_{\hbox{\scriptsize max}})$ occur,
see figure \ref{Fig:Q_ilc_KQ85}.
This might be too large for cosmological applications.
It is also checked that refining the resolution from
$N_{\hbox{\scriptsize side}} = 16$, which is used in the above computations,
to $N_{\hbox{\scriptsize side}} = 32$ does not improve the reconstruction.

Furthermore, the behaviour of the temperature 2-point
correlation function $C(\vartheta)$ is analysed
with respect to the various reconstructed maps,
i.\,e.\ with and without additional smoothing,
for the ILC (7yr) map and for CMB simulations of the 
$\Lambda$CDM concordance model.
The correlation function $C(\vartheta)$ is computed from
the full ILC map, from the data outside the KQ85 mask,
and from the reconstructed ILC maps
which depend on $l_{\hbox{\scriptsize max}}$.
In order to get a good approximation of the
correlation function $C(\vartheta)$ on large angular scales,
the multipoles up to $l_{\hbox{\scriptsize max}} \simeq 10$ are needed.
The application of the large KQ75 mask is therefore excluded.
The errors of the reconstructed 2-point correlation function
are estimated by using 1000 CMB simulations of the $\Lambda$CDM model.
Since this error estimation is based on simulations
which contain only the pure CMB signal,
no further errors are taken into account 
e.\,g.\ resulting from residual foreground and detector noise.
Therefore, the error estimation of the correlation function
of the ILC map gives only a lower bound of the errors
expected in a genuine application.

As discussed above the reconstructed ILC map with additional smoothing 
of 600 arcmin uses information from inside the mask. 
This information transfer leads in turn to underestimated errors
in the 2-point correlation function.
Furthermore, the similarity of $C(\vartheta)$ computed from the full ILC map
and from the reconstructed map is due to this information transfer
and not the achievement of the reconstruction algorithm.

Without additional smoothing
the ILC correlation function $C(\vartheta)$,
computed from the reconstructed map, does not match
the one of the full map better than the one of the masked map
within the estimated errors.
This is caused by the large errors
which increase with $l_{\hbox{\scriptsize max}}$.
For $10 \lesssim l_{\hbox{\scriptsize max}} \lesssim 12$
the reconstructed 2-point correlation function is uncertain
by 100$\mu\hbox{K}^2$ and for even larger values
of $l_{\hbox{\scriptsize max}}$ useless as revealed by
figure \ref{Fig:correlation_function_ilc_KQ85_FWHM_000arcmin_nside_16}.

This uncertainty is also reflected in the $S(60^{\circ})$ statistic
which integrates the power of the 2-point correlation function
on angular scales larger than $60^{\circ}$.
The $S(60^{\circ})$ statistic of the reconstructed ILC map 
does not converge to a stable value
for $l_{\hbox{\scriptsize max}} \gtrsim 6$
as it is the case for the full and for the masked ILC map
as displayed in figure 
\ref{Fig:s_statistik_1000lcdm_und_ilc_KQ85_FWHM_000arcmin_nside_16}a.
In addition, the mean value of the $S(60^{\circ})$ statistic,
calculated from 1000 reconstructed CMB sky maps,
reveals the instability of the algorithm
for $l_{\hbox{\scriptsize max}} \gtrsim 12$,
where the power strongly increases, see the dashed curve in figure
\ref{Fig:s_statistik_1000lcdm_und_ilc_KQ85_FWHM_000arcmin_nside_16}b.
This demonstrates the inability to reconstruct the temperature correlations.

The conclusion of this paper is that by using a realistic mask for
the WMAP data the reconstruction algorithm (\ref{Eq:ar_by_A})
does not work well enough to obtain a prediction of
the 2-point correlation function. 
For this reason a cosmological analysis of the WMAP data 
should use only data outside a mask,
e\,g.\ the KQ85 (7yr) or KQ75 (7yr) mask.
The Planck satellite measures the sky at more different 
frequencies than the WMAP satellite.
This should allow a more secure reduction of the foreground in the CMB map.
It is expected that the masked region in the Planck data is
significantly smaller.
It might be small enough in order to allow an acceptable reconstruction
of the CMB within the mask which could provide 
a 2-point correlation function useable at large scales.



\section*{Acknowledgments}

We would like to thank the Deutsche Forschungsgemeinschaft
for financial support.
HEALPix (healpix.jpl.nasa.gov)
\citep{Gorski_Hivon_Banday_Wandelt_Hansen_Reinecke_Bartelmann_2005}
and the WMAP data from the LAMBDA website (lambda.gsfc.nasa.gov)
were used in this work.
The computations are carried out on the Baden-W\"urttemberg grid (bwGRiD).


\bibliography{../bib_astro}
\bibliographystyle{apalike}

\label{lastpage}

\end{document}